\documentclass[
reprint,
superscriptaddress,
 amsmath,amssymb,
 aip,
 jcp,
]{revtex4-1}
\usepackage{graphicx}
\usepackage{subfigure}
\usepackage{dcolumn}
\usepackage{bm}
\usepackage{multirow}

\newcommand{\bea}{\begin{eqnarray}}
\newcommand{\eea}{\end{eqnarray}}

\newcommand{\Eq}[1]{Eq.\,(\ref{#1})}

\begin{document}


%
%
\title{ Calculating vibrational spectra with sum of product basis
 functions without storing full-dimensional vectors or matrices}
\author{Arnaud Leclerc}
\email{Arnaud.Leclerc@univ-lorraine.fr}
\affiliation{Chemistry Department, Queen's University, Kingston, Ontario K7L 3N6, Canada}
\affiliation{Universit\'e de Lorraine, UMR CNRS 7565 SRSMC, Th\'eorie-Mod\'elisation-Simulation, \\1, boulevard Arago 57070 Metz, France}
\author{Tucker Carrington}
\email{Tucker.Carrington@queensu.ca}
\affiliation{Chemistry Department, Queen's University, Kingston, Ontario K7L 3N6, Canada}


\begin{abstract}

We propose an iterative method for computing vibrational spectra that significantly reduces the memory cost
of calculations.   It uses a direct product primitive basis, but does not require storing vectors with as many 
components as there are product basis functions.  Wavefunctions are represented in a basis each of whose 
functions is a sum of products (SOP) and the factorizable structure of the Hamiltonian is exploited.   
If the factors of the SOP basis functions are properly chosen, wavefunctions are linear combinations of a 
small number of SOP basis functions.   The SOP basis functions are generated using a shifted block power method.  
The factors are refined with a  rank reduction algorithm to cap the number of terms in a SOP basis function. 
The ideas are tested  on a 20-D model Hamiltonian and a realistic 
 CH$_3$CN  (12 dimensional)  potential.  For the 20-D problem, to 
 use  a standard direct product iterative approach one would need to store vectors with about
$10^{20}$ components and would hence require about   $8 \times 10^{11}$ GB.   With the approach of this paper only 1    GB
of memory is necessary.   Results for  CH$_3$CN agree well with those of a previous calculation on the same 
potential.

\end{abstract}
\maketitle

\section{Introduction}

The most general and systematic way of solving the time-independent Schr\"odinger equation to compute 
vibrational bound states, and hence a vibrational spectrum,  requires computing eigenvalues of a 
basis representation of the Hamiltonian operator.  When standard methods of ``direct'' linear algebra are used
to diagonalize the Hamiltonian matrix the memory cost of the calculation scales as $N^2$, where $N$ is the size of  the 
matrix and the number of basis functions.  Diagonalization can be avoided  by using an iterative eigensolver to 
compute  the eigenvalues of interest.   The Lanczos\cite{lanczos,cullum} and Filter Diagonalization 
methods\cite{neuhauser1990,mandel,cg,shi} are popular iterative options.
Iterative approaches require only the evaluation of matrix-vector products.  If it is not 
possible to do   matrix-vector products without keeping the Hamiltonian matrix in memory then the memory cost 
of iterative methods also scales as    $N^2$.  Fortunately, one can often exploit either structure of the 
basis (and the Hamiltonian operator) or sparsity of the Hamiltonian matrix to evaluate matrix-vector products
without storing (and sometimes without computing elements of) the Hamiltonian matrix.  
\cite{hd,mjbdirect,corey}
In both cases, the memory cost of a product basis iterative calculation  scales as
$N=n^D$,
 where $n$ is a representative number of basis  functions for a single coordinate and $D$ is the number of dimensions,
which is  the size of a vector (at least two vectors must be retained in memory).  
Exploiting the structure of  a   product basis also makes it possible to evaluate matrix-vector products efficiently
(at a cost that scales, regardless of the complexity of the potential, as $n^{D+1}$). 
                  \cite{mjbdirect,corey,lef,lef2,pranab,cgss,jc,cghooh,nno,gustc2h4,dvrrev} 

 The combination of 
iterative algorithms and product basis sets reduces the memory cost to $n^D$,  which is obviously much less 
than $n^{2D}$, required to store the Hamiltonian matrix in the product basis.   However, even  $n^D$ is 
very large if $D > 6$.  If $D=12$ and $n=10$ then for  a single vector one needs  $ \sim 8000$ GB of memory.  
This is a manifestation of the ``curse of dimensionality''. 
To further reduce the memory cost one option is to   use a better basis.  
 It is not hard to devise a better basis, what $is$
tricky is finding a basis that has enough structure to make it possible to  efficiently
evaluate matrix-vector products. If matrix-vector products are not evaluated efficiently, 
reducing the number of basis functions
required to represent the wavefunctions of interest can  (significantly) $increase$ the CPU cost of 
the calculation.   There are four popular ways of reducing basis size.  First, prune a standard  product basis by retaining
 only some of the functions.  \cite{joel,benoit,rau,sd1,sd2,nick,colbert,yu,wyatt}  Second, use contracted basis functions obtained by 
solving reduced-dimension eigenproblems.
\cite{cont1,cont2,cont3,cont4,cont5,cont6,cont7,cont8}
  Third, optimize 1D functions with the multi-configuration time dependent
Hartree method (MCTDH).\cite{mctdhbook,mctdhrev}
Fourth, use basis functions localized in the classically allowed region of phase space.   
\cite{davis,bill,halverson2012,tannor}
Some pruned bases are compatible with efficient matrix-vector products, for both problems with simple potentials
(requiring no quadrature)\cite{const} and with general potentials (for which quadrature is necessary)
\cite{gustc2h4,gustnonprod,gustbetter,lauvergnat2013}.  Several ideas have been proposed for evaluating matrix-vector products with contracted bases.
\cite{mjbcont,xgwcont,rich,claude}  Because MCTDH uses a direct product basis it is straightforward to evaluate
matrix-vector products at a cost that scales as $n^{D+1}$ (with $n$ the number of single-particle functions).
To date no one has attempted to use
iterative methods in conjunction with  phase space localized bases.

In this paper we propose an iterative method, for computing spectra, that significantly reduces  
 the memory cost of calculations.   We use a direct product basis (although the ideas would also work with a 
pruned basis).  
 To represent a wavefunction, all previous product-basis iterative methods   store    
 $n^D$ coefficients.
     Our new approach is motivated by the realization that,  in some cases,  the  $n^D$
coefficients, used to represent a function, 
can be computed from a much smaller set of numbers.   For example, 
 a product of functions of a single variable, $\phi_1(q_1) \phi_2(q_2) \cdots \phi_D(q_D)$, can be
represented as 
\begin{equation*}
\sum_{i_1=1}^{n} f^{(1)}_{i_1} \theta^1_{i_1}(q_1) 
\sum_{i_2=1}^{n} f^{(2)}_{i_2} \theta^2_{i_2}(q_2)  \cdots
 \sum_{i_D=1}^{n}   f^{(D)}_{i_D} \theta^D_{i_D}(q_D) 
\end{equation*}
 and it is only necessary to store $D n$ numbers.   Obviously, for a real problem the wavefunction is not
a  product of functions of a single variable, but it should be possible to represent many 
wavefunctions as sums of products of  functions of a single variable. 
 If, for one wavefunction, 
 $R$ terms are required, one must
store $ R D n$ numbers.   This may be much less than $n^D$.   When $n=10$ and  $D=12$,  
 $ R D n <  n^D$ if $R <  8 \times 10^9$.    
For many molecules it is surely  possible to find a sum of products (SOP) representation of wavefunctions with a value 
of $R$ small enough that it is worth exploiting the SOP structure  to reduce the memory cost.

We develop a method using SOP basis functions to find eigenpairs of a SOP operator in section \ref{theo}.  The memory cost scales as $nRD$,  which is the 
memory required to  store one SOP basis function,     where $R$ is the required number
of terms in a SOP basis function. 
   The key idea is to use basis functions that are sums of products of 
optimized factors. 
Basis functions are determined, 
from matrix-vector products evaluated by doing 1-D operations, 
 by applying the Hamiltonian to other SOP functions.  
 The number of terms in the basis functions is controlled by a reduction procedure. 
The reduction is a crucial part of the method we propose.   
In section \ref{models}, the  method  is  tested on multidimensional coupled oscillator models with D as large as    20.  The lowest transitions of acetonitrile, 
 CH$_3$CN,    (a 12-D problem) are computed and compared with results of    Avila et al\cite{avila2011} 
in    section  \ref{ch3cn}.

\section{Sum of products (SOP) eigensolver\label{theo} }

\subsection{SOP basis functions and CP format representation}

Our goal is to calculate eigenstates of a  Hamiltonian operator 
by representing it in an efficient  SOP basis.   
We define a primitive product basis using 1-D functions $ \theta^j_{i_j} (q_j) $ with $i_j=1,\dots,n_j$ for each coordinate $q_j$.  The primitive basis is unusably large.  
An  SOP basis function,   $  \Psi_k (q_1, \dots, q_D)  $,    can be expanded in the primitive  basis as
\begin{equation}
\Psi_k (q_1, \dots, q_D) \simeq \sum_{i_1=1}^{n_1} \dots \sum_{i_D=1}^{n_D} F_{i_1 i_2\dots i_D}
\prod_{j=1}^D \theta_{i_j}^j (q_j)~.
\label{wavefunction}
\end{equation}
 For SOP basis functions, 
\begin{equation}
F_{i_1 i_2 \dots i_D} = \sum_{\ell=1}^R  \prod_{j=1}^D f^{(\ell,j)}_{i_j} ~,
\label{sop}
\end{equation}
where $f^{(\ell,j)}$ is a one-dimensional vector associated with  the $\ell$-th term and coordinate $j$,
 and there is no need to work
 explicitly with $  F_{i_1 i_2\dots i_D} $,  which is a $D$-dimensional
 tensor with $~n^D$ components,  
For example,  if $D=2$,  a SOP basis function with two terms has the form
\bea  
&& c^1(q_1) g^1(q_2) + c^2(q_1) g^2(q_2) \nonumber \\
&=&  \sum_{i_1}  f^{(1,1)}_{i_1} \theta^1_{i_1}(q_1)   \sum_{i_2}  f^{(1,2)}_{i_2} \theta^2_{i_2}(q_2)
 \nonumber \\
&+& 
\sum_{i_1}  f^{(2,1)}_{i_1} \theta^1_{i_1}(q_1)   \sum_{i_2}  f^{(2,2)}_{i_2} \theta^2_{i_2}(q_2)
 \nonumber \\
&=&  \sum_{\ell=1}^2  \sum_{i_1}     \sum_{i_2}      f^{(\ell,1)}_{i_1}  f^{(\ell,2)}_{i_2}   \theta^1_{i_1}(q_1)  \theta^2_{i_2}(q_2).
\eea
 $F_{i_1 i_2 \dots i_D} = \sum_{\ell=1}^R  \prod_{j=1}^D f^{(\ell,j)}_{i_j}$ represents the  
 function in the primitive 
$
\prod_{j=1}^D \theta^j_{i_j} (q_j)$ basis.  
This SOP format for multidimensional functions is known as the canonical polyadic (CP) decomposition for tensors in the applied mathematics literature  \cite{zhang2001,beylkin2002,beylkin2005} (also called parallel factor decomposition or separated representation).
Truncating the sum at a given rank $R$ gives a reduced rank approximation for $F$. 
The CP format has been successfully applied to the calculation of many-body electronic integrals and
 wavefunctions.\cite{benedikt2011,bischoff2011,bischoff2012,hohenstein2012,parrish2012,hohenstein2012_2} 
Because the factors in the terms are not chosen from a pre-determined set,  our basis functions are in the 
CP format.

There are other  reduced (compressed) tensor formats which could be used 
 to   compactly  represent  $  F_{i_1 i_2\dots i_D} $.
  The most familiar compression of this type, for a 2-D problem, 
 is the singular value decomposition.   
For $D>2$, different decompositions exist.\cite{kolda2009}
In the Tucker format
\cite{hitchcock1927,tucker1966,kolda2009},
$
F_{i_1 i_2\dots i_D} = \sum_{\ell_1=1}^{L_1} \dots \sum_{\ell_D=1}^{L_D}  K_{\ell_1 \ell_2 \dots \ell_D} 
\prod_{j=1}^D  a ^{(j)}_{i_j \ell_j}.
$
where $K$ is called the core tensor  and  $L_j < n_j \quad \forall j=1\dots 
D$ and the $a^{(j)}$ are $n_j \times L_j$ matrices. 
This format is equivalent to the one used by MCTDH.\cite{mctdhbook}
The Hierarchical Tucker format\cite{hackbusch2009,grasedyck2010,kressner2011} (of which the tensor train
 format\cite{oseledets2011}  is a particular case)    is   a compromise  between the Tucker format and the CP 
 format.  
 It was first introduced by developers of MCTDH \cite{wang2003}.

In this article we propose a procedure for making SOP basis functions in the form of Eq. \eqref{sop}. 
How do we make the basis functions?   We shall begin with a function   having  one term (i.e. with  rank 1) that is
 obtained from 
   $ F_{i_1 i_2 \dots i_D}     =  \prod_{j=1}^D f^{(1,j)}_{i_j}$ 
with some random $ f^{(1,j)}_{i_j}$ 
and  obtain basis functions  (see the next subsection) by applying the Hamiltonian operator.
  Throughout this paper we shall assume that 
 the Hamiltonian  is   also  a sum of products,
\begin{equation}
H(q_1,\dots,q_D) = \sum_{k=1}^T \prod_{j=1}^D  h_{k j} (q_j),
\label{hamiltonian}
\end{equation}
where $h_{kj}$ is a one-dimensional operator acting in a Hilbert space associated with   coordinate 
$q_j$.
Kinetic energy operators (KEOs) almost always have this form.  If the potential is not in SOP form it can be 
massaged into SOP form by using, for example, potfit \cite{mctdhbook,mctdhrev}, multigrid potfit\cite{multipot}, or neural network methods \cite{nnsop,nnsop2,nnsop3,browna}.

\subsection{Shifted power  method 
 \label{method}}

In this subsection we explain how SOP basis functions are made by applying the Hamiltonian.  
In  the $\prod_{j=1}^D \theta_{i_j}^j (q_j)$ basis the SOP basis functions are represented by the 
 $F_{i_1 i_2 \dots i_D}$  coefficients  in \Eq{sop}.  We use the  power method,
  the simplest iterative method\cite{strang1986,saad2011}, to determine 
 the  $ f^{(\ell,j)}_{i_j}  $. 
 Let ${\bf F }^{(0)}$ be a random start 
 vector  of the form of \Eq{sop} 
and ${\bf V}_{E_{\text{max}}}$ be the eigenvector associated with  the  eigenvalue,  $E_{\text{max}}$,  whose absolute value is largest.  Throughout this paper we shall assume that  the minimum potential energy is zero 
and therefore  that all  eigenvalues of {$\bf{H}$}, the finite matrix representing the Hamiltonian in the 
primitive $\prod_{j=1}^D \theta_{i_j}^j (q_j)$ basis,  are positive.  In this case,   
  $E_{\text{max}}$ is simply  the largest 
eigenvalue.
Assuming  $({\bf F}^{(0) })^T  {\bf V}_{E_{\text{max}}} \neq 0$, 
\begin{equation}
\lim_{N_{\text{pow}}\to \infty} {\bf H}^{N_{\text{pow}}} {\bf F}^{(0)} \rightarrow {\bf V}_{E_{\text{max}}}~.
\end{equation}
When   $N_{\text{pow}}  $ is large, 
  ${\bf F}^{(N_{\text{pow}})} = {\bf H} ^{N_{\text{pow}}} {\bf F}^{(0)}   $   approaches 
the eigenvector of  ${\bf H}$ with the largest eigenvalue.
If the Hamiltonian is a SOP (\Eq{hamiltonian}) then 
 ${\bf F}^{(N_{\text{pow}})} $ has the form of \Eq{sop}.  
The convergence of the power method is known to be slow and to depend on  gaps between eigenvalues close to
  $E_{\text{max}}$  and  $E_{\text{max}}$. \cite{strang1986,saad2011}  
The error is 
approximately    proportional to 
$ (E_{sl}/ E_{\text{max}}) ^{N_{\text{pow}}}   $, where $ E_{sl}  $  is the second largest eigenvalue.  
We could use the  ${\bf F}^{(N_{\text{pow}})}$ sequence  to compute  
 the largest eigenvalue of  ${\bf H}$.  Each  ${\bf F}^{(N_{\text{pow}})}$ 
has  the form of 
 \Eq{sop} and hence its  storage requires little memory.  However, 
 we do not want the largest eigenvalue and we wish  to compute more than 
one eigenvalue.   From a reasonable estimate of  $E_{\text{max}}$ one can obtain the smallest eigenvalue of  {$\bf{H}$},
from the linearly shifted operator
\begin{equation}
\widetilde{H} = H- \sigma \openone   ~,~~ \sigma =   E_{\text{max}}   ~.
\end{equation}
When several eigenstates are desired, one uses  a block method  which begins with a set of $B$ random start
    vectors. Alternating successive 
applications of $\widetilde{\bf H}$ with a modified Gram-Schmidt orthogonalization, we   obtains a set of vectors,
each of the form of \Eq{sop},
 which converges to the eigenvectors associated with the lowest eigenvalues of  {$\bf{H}$}.  
These are  SOP basis vectors.    
Orthogonalization 
requires adding  vectors which is done by concatenation.  This  increases the rank.   
  It  is not necessary to 
orthogonalize after every application of   $\widetilde{ \bf H}$, instead the 
 orthogonalization can be done
 only after each set of  $N_{\text{ortho}}$  matrix-vector products. 
The convergence of the shifted block method is somewhat less  slow  than the convergence of the simple 
power method  and  now depends on the gaps between the $B$ smallest    eigenvalues and the 
($B+1$)th smallest  eigenvalue  of  {$\bf{H}$}.  Gaps between the $B$ smallest    eigenvalues  play no role.  
 Degeneracies, within the block,  cause no problems.
Rather than shifting with $E_{\text{max}}$, it is better to shift with a value slightly larger than the average
of  $E_{\text{max}}$ and the ($B+1$)th  eigenvalue   of {$\bf{H}$}. \cite{parlett}
In practice we use, 
\begin{equation}
\sigma_{\text{opt}} = \frac{    E_{\text{max}}    + E_>}{2}~,
\label{idealshift}
\end{equation}
where $   E_>   $ is an upper bound for the $(B+1)$th eigenvalue of  {$\bf{H}$}.  
The desired eigenvalues of    ${\bf H}$  correspond to the largest eigenvalues of    $\widetilde{ \bf H}$.
The algorithm can be first applied to   {$\bf{H}$}    with $\sigma = 0$ and $B=1$ to calculate $E_{\text{max}}$.
 $   E_>   $ is obtained by running a few iterations of the algorithm with  $\sigma =  E_{\text{max}}  $ and a block size $B+1$. 
The algorithm also works with    the non-optimal shift $\sigma = E_{\text{max}}$.
The vectors obtained by successively applying  $(\widetilde{ \bf H})$ to a set of $B$ start vectors and
orthogonalizing  will, if 
  $N_{\text{pow}}$  is large enough, approach  the 
  matrix   of   eigenvectors  ${\bf V} = ( {\bf V}_1 \dots {\bf V}_B )$.   We denote these vectors
\begin{equation}
\bm{\mathcal{F}} = ( {\bf F}_1^{(N_{\text{pow}})}  \dots  {\bf F}_B^{(N_{\text{pow}})})~.  
\label{basis}
\end{equation}
 $\bm{\mathcal{F}}$ can also be used as a basis for representing 
 ${\bf H}$, to obtain more accurate eigenvalues and eigenvectors.  
$  \bm{\mathcal{F}}$ is   our SOP basis.  
Even if  $N_{\text{pow}}$  is not large enough to ensure that {$\bm{\mathcal{F}}$} is a set of eigenvectors, 
 the subspace spanned by the  
{$\bm{\mathcal{F}}$}
set may  be sufficient to obtain good approximations for  the smallest 
eigenpairs by projecting into the space, i.e., by computing eigenpairs of the 
generalized eigenvalue problem, {$\bf{ H^{(\bm{\mathcal{F}})} U =  S U E}$}, where 
 ${\bf H ^ {(\bm{\mathcal{F}})}} = \bm{\mathcal{F}}^T {\bf H}  \bm{\mathcal{F}}$  and  ${{\bf S}}  =
 \bm{\mathcal{F}}^T  \bm{ \mathcal{F}}$. 
A simple eigenvalue problem would be sufficient to obtain the eigenvectors if the $\bm{ \mathcal{F}}$ basis set 
were always perfectly orthogonal. 
However  residual non-orthogonality is present, due to
 a reduction 
(compression) step that must be introduced into the algorithm (see section \ref{reduction_section}).
This explains why a generalized eigenvalue problem is  be used.
${\bf S}$ is computed by doing one-dimensional operations,\cite{beylkin2005} 
(${\bf F}$ and ${\bf F'}$ being  two vectors of the $\bm{ \mathcal{F}}$ set),  
\begin{equation}
{\bf F}^T {\bf    F'}  
 = \sum_{\ell=1}^R \sum_{\ell'=1}^{R'}   \prod_{j=1}^D ({\bf f}^{(\ell,j) })^T  {\bf  f'}^{(\ell',j)} 
\label{scalarproduct}
\end{equation}  
with  
$
 ({\bf f}^{(\ell,j) } )^T  {\bf  f'}^{(\ell',j)} 
 = \sum_{i_j=1} ^{n_j}
f^{(\ell,j)}_{i_j} f'^{(\ell',j)}_{i_j} $.
${\bf H}^{(\bm{\mathcal{F}})} $ matrix elements are computed similarly, from 1-D operations
 (Eq. \eqref{hpsiproduct} of section \ref{reduction_section} followed by a scalar product using an approach similar to that of  Eq. \eqref{scalarproduct}).
  It is advantageous to restart,  every 
 $N_{\text{diag}}$ iterations,   with approximate eigenvectors of {$\bf{H}$} that are columns of
${\bm{\mathcal{F}}}{ \bf{U}}$.
  In our programs
  $N_{\text{diag}}$ can be equal to or a 
multiple of $N_{\text{ortho}}$ (see section \ref{models} and \ref{ch3cn}).
The approximate eigenvectors used to re-start  are sums of $B$ different
 ${\bf F}_k^{(N_{\text{pow}})} $ vectors
and are obtained by  concatenating, which increases the rank to $BR$  
if every ${\bf F}_k^{(N_{\text{pow}})}$ has rank $R$.

In this section,  we specify the SOP basis functions.   
Thus far it appears that they    are 
obtained from 
the vectors of 
\Eq{basis}.     This, however, is not practical.  Applying 
  {$\bf{H}$} to a vector (and even re-starting  and orthogonalizing)  increases the rank of the vectors (the number of
products in the sum) and causes the memory cost to explode.   In the next subsection we outline how to 
obviate this problem by reducing the rank of the SOP basis functions.    
  The combination of an iterative  algorithm for making a SOP basis and rank reduction 
will only work if the basis vectors generated by the iterative algorithm converge to low-rank vectors.  If they
do, reducing the rank of the vectors will cause little error.  If they do not, and even if eigenvectors are 
 low-rank and linear combinations of basis vectors  
 generated by the iterative algorithm,   
 reducing the rank of the vectors will cause significant  error.  
We expect eigenvectors of  {$\bf{H}$}  to be low rank and therefore expect it to be possible to reduce the rank of
vectors generated by the shifted 
power method, which approach eigenvectors. 
The slow convergence of the power method is thus compensated by the advantage of being able to work with 
low rank vectors.   

\subsection{{\bf H} application and rank reduction \label{reduction_section}}

The key step in  the block power method  is the application of 
  {$\bf{H}$} to a  vector  ${\bf F}$  to obtain a new vector    ${\bf F}'$.  
With $T$ terms in  {$\bf{H}$},  
  the rank of   ${\bf F'}$  is a factor of $T$ larger than the rank of  ${\bf F}$.
All vectors are represented as  
\begin{equation}
F_{i_1 i_2 \dots i_D} = \sum_{\ell=1}^R s_{\ell} \prod_{j=1}^D \tilde{f}^{(\ell,j)}_{i_j} \text{ with } 
\sum_{i_j}^{n_j} \vert \tilde{f}^{(\ell,j)}_{i_j} \vert^2 = 1~,
\end{equation}
where, for each term ($\ell$) and each coordinate ($j$), $\tilde{f}^{(\ell,j)}_{i_j}$ is a normalized 1-D vector, 
 $ s_{\ell} $   is a normalization  coefficient, and $n_j$ is the number of basis functions for coordinate
$j$.  
Using normalized 1-D vectors allows us to order the different terms in the expansion.
This is useful  for
identifying   dominant terms in the sum.
  {$\bf{H}$} can be applied to ${\bf F} $ by evaluating 1-D matrix-vector products with 
matrix representations of  1-D operators ${\bf h}_{kj}$ in the $\theta_{i_j}^j$ basis, i.e.
$({\bf h}_{kj})_{i_j,i_j'} = \langle \theta_{i_j}^j \vert  h_{k j} \vert \theta_{i_j'}^{j} \rangle$,
and 1-D vectors  $\tilde{f}^{(\ell,j)}_{i_j}$,
\bea
&& ({\bf F'})_{i'_1 i'_2 \cdots i'_D}  =  ({\bf H F})_{i'_1\dots i'_D} \nonumber \\
&&	 =  \sum_{i_1, i_2, \cdots, i_D}   
			\sum_{k=1}^T     \prod_{j'=1}^D  ({\bf h}_{kj'})_{i'_{j'} i_{j'}}
			 \sum_{\ell=1}^{R} \prod_{j=1}^D    s_{\ell} \tilde{f}^{(\ell,j)}_{i_j} \nonumber \\
&& =  \sum_{k=1}^T   \sum_{\ell=1}^{R} \prod_{j=1}^D 
\sum_{i_j}    
  ({\bf h}_{kj})_{i'_{j} i_{j}}
    s_{\ell} \tilde{f}^{(\ell,j)}_{i_j}.
\label{hpsiproduct}
\eea
   Applying   {$\bf{H}$} to    ${\bf F}$, with  $R$ terms,
yields a vector with $RT$ terms.    Owing to the fact that everything is done with 1-D matrix-vector products,
generating the vector    ${\bf F}'$       is inexpensive.

If the rank were not reduced after each matrix-vector product, the rank of a vector obtained by applying  {$\bf{H}$} \;  $P$
times  to a   start vector with  $R_0$ terms would be 
$
T^P R_0 $.
If $T$ and/or $P$ is large, one would  need more, and not less, memory to store  the vector than would be 
required to store  $n^D$ components.    Table \ref{Plimit} shows, for $n=10$, 
%
%
the maximum value of $P$ for which 
less memory is needed to store a vector obtained by applying 
 {$\bf{H}$} \; $P$
times  to a 
 start vector with  rank one ($R_0=1$).  
This table clearly reveals that rank reduction is  imperative.  
\begin{table}[ht]
\caption{Maximum number of products {$\bf{H}$}  {$\bf{F}$}
 before losing the memory advantage of the CP format if 
$H$ has $T$ terms, in $D$ dimensions.}
\begin{tabular}{llllll}
\hline
\hline
$T \setminus D$ & 3 & 6 & 12 & 20 & 30\\
 \hline
15 & 2 & 5 & 10 & 17 & 25\\
30 & 2 & 4 & 8 & 13 & 20\\
100 & 1 & 3 & 6 & 10 & 15\\
200 & 1 & 2 & 5 & 8 & 13 \\
400 & 1 & 2 & 4 & 7 & 11 \\
\hline
\hline
\end{tabular}
\label{Plimit}
\end{table}

What algorithm is used to    reduce the rank and by how much is the rank reduced?  To reduce the rank, we replace 

\bea
F^{\text{old}}_{ i_1 i_2 \dots i_D}  &=& \sum_{\ell=1}^{R_{\text{old}}}  { ^{\text{old}}s_{\ell} }\prod_{j=1}^D  {^{\text{old}}\tilde{f}}^{(\ell,j)}_{i_{j}}
 \nonumber \\
&\Longrightarrow &
F^{\text{new}}_{ i_1 i_2 \dots i_D} 
= \sum_{\ell=1}^{R_{\text{new}}} { }^{\text{new}}  s_{\ell} \prod_{j=1}^D { }^{\text{new}}\tilde{f}^{(\ell,j)}_{ i_j}~,
\eea
 where ${R_{\text{new}}} < R_{\text{old}}$
and choose $ ^{\text{new}}\tilde{f}^{(\ell,j)}_{ i_j}$ to minimize $
\parallel {\bf F}^{\text{new}} - {\bf F}^{\text{old}}\parallel
$.   
Making this replacement changes a  vector generated by the power method, but because energy levels are computed
by projecting into the space spanned by  $
\bm{\mathcal{F}}$,   numerically exact results can still be obtained.  If $  {R_{\text{new}}} \sim {R_{\text{old}}}  $,
 $ \parallel {\bf F}^{\text{new}} - {\bf F}^{\text{old}}\parallel$ is small but the memory cost is large.  
One might choose  ${R_{\text{new}}}$, for each reduction,  so that   $
\parallel {\bf F}^{\text{new}} - {\bf F}^{\text{old}}\parallel$ is less than some threshold. Instead, we use the same 
 ${R_{\text{new}}}$ for all reductions and choose a value  small enough that the memory cost is much less than 
the cost of storing $n^D$ components but large enough that good results are obtained from a relatively small 
value of $N_{\text{pow}}$.  Rank reduction is motivated by the realization that when the Hamiltonian is separable, i.e.
$H(q_1,\dots,q_D) = h_1 (q_1) + h_2 (q_2) + \dots + h_D (q_D) $, the 
wavefunctions are all of rank one and when coupling is not huge the rank of wavefunctions is small (it is important 
to understand that the rank of a wavefunction is not the same as the number of 
 $\prod_{j=1}^D \theta_{i_j}^j (q_j)$
basis functions which  contribute 
to it).  In general, the stronger the coupling, the larger the required value of   ${R_{\text{new}}}$.  Note that 
wavefunctions are represented as linear combinations of basis functions with rank  ${R_{\text{new}}}$ and may therefore have
rank larger than   ${R_{\text{new}}}$.

We use  an alternating least squares (ALS) algorithm  described in 
Ref. \onlinecite{beylkin2005} to determine the $
^{\text{new}}\tilde{f}^{(\ell,j)}_{ i_j}$ by minimizing  $
\parallel {\bf F}^{\text{new}} - {\bf F}^{\text{old}} \parallel  $.  
The reduction algorithm needs  start values for  $
{^{\text{new}}\tilde{f}}^{(\ell,j)}_{ i_j} $.   We use the   ${R_{\text{new}}}$ 
terms in $
 {\bf F}^{\text{old}} $ 
 with the largest  $s_{\ell}$ coefficients.   
 Another possibility is to use random start vectors.
For all $l$ values   $
^{\text{new}}\tilde{f}^{(\ell,j)}_{ i_j}$  factors are varied, for a single $j=k$, keeping    
 all the other factors  $\tilde{f}^{(\ell,j)}_{i_j} \quad \forall j\neq k$ fixed
 (and then the vectors are normalized by changing     $s_{\ell}$). 
For each coordinate  we solve
\begin{equation}
\frac{\partial \parallel {\bf F}^{\text{new}} - {\bf F}^{\text{old}} \parallel}{\partial ^{\text{new}}\tilde{f}^{(\ell,k)}_{ i_k}} = 0 \quad \forall \ell , \quad \forall i_k.
\end{equation}
For a single coordinate $k$, this requires solving   
linear systems  with an  $(R_{\text{new}} \times R_{\text{new}})$ matrix whose elements are
\begin{equation}
B ( \hat{\ell}, \tilde{\ell} ) = \prod_{\substack{j=1 \\ j \neq k}} ^{D} 							
	\left(  {}^{\text{new}}\tilde{\bf f}^{ (\hat{\ell}, j)  } \right)^T  
 {}^{\text{new}}\tilde{\bf f}^{ (\tilde{\ell}, j) }
\end{equation}
and with $n_k$ different right-hand-sides, the  $i_k$th of which is   
\begin{equation}
d_{i_k} (\hat{\ell}) =  \sum_{\ell =1}^{R_{\text{old}}}
						{ }^{\text{old}} s_l \;  ^{\text{old}} \tilde{f}^{({\ell}, k)}_{i_k} 
						\prod_{\substack{j=1 \\ j \neq k}} ^{D} 
						\left( {}^{\text{old}} \tilde{\bf f} ^{ ( \ell, j ) } \right) ^T  \;
 {}^{\text{new}}\tilde{\bf f}^{ (\hat{\ell}, j) } ~.
\end{equation}
$   {}^{\text{new}} \tilde{ f}_{i_j}^{ (\tilde{\ell}, j) } $ is obtained by solving, 
\begin{equation}
   \sum_{\tilde{\ell}}       B (\hat{\ell}, \tilde{\ell} )		
	  {}^{\text{new}} \tilde{ f}_{i_j}^{ (\tilde{\ell}, j) } ~ =  ~ d_{i_j} (\hat{\ell})  ~.
\end{equation}
Ill-conditioning is avoided with a penalty term as described in Ref. \onlinecite{beylkin2005}. 
   Repeating this  for all  $D$ coordinates  constitutes 
one ALS iteration,  with a computational cost of 
\begin{equation}
\mathcal{O}\left(D (R_{\text{new}}^3+ n (R_{\text{new}}^2  + R_{\text{new}}R_{\text{old}}))\right),
\end{equation}
where  $n$ is a representative value of  $n_j$.  
$\mathcal{O} (D n R_{\text{new}}^2)$ is the cost of making 
the ${\bf B}$ matrices (${\bf B}$ matrices for successive coordinates are made by updating\cite{beylkin2005}),
$ \mathcal{O} (D n R_{\text{new}} R_{\text{old}}) $ is the cost of computing the  right-hand-sides, 
and $\mathcal{O} (D R_{\text{new}} ^3)$ is the cost of solving the linear systems.

One could iterate the ALS algorithm until $
\parallel {\bf F}^{\text{new}} - {\bf F}^{\text{old}} \parallel  $ is less than some pre-determined threshold.   
Instead, we fix the number of ALS iterations, $N_{ALS}$,  
 on the basis of preliminary tests.     
We do this because computing 
 $
\parallel {\bf F}^{\text{new}} - {\bf F}^{\text{old}} \parallel  $ is costly 
as it requires the calculation of many scalar products with vectors of rank $(R_{\text{new}}+R_{\text{old}})$, which scales as $\mathcal{O}((R_{\text{new}}+R_{\text{old}})^2)$. 
If  ${\bf F}^{\text{new}} $ determined by fixing the number of ALS iterations is not a good approximation
to   ${\bf F}^{\text{old}} $ then we alter the vector obtained from the block power method more than we would like, however,
this changes only a basis vector and does not preclude computing accurate energy levels.
More effective or more efficient reduction algorithms exist, such as the Newton method of 
Ref. \onlinecite{zhang2001,chinnamsetty2007}  and the  conjugate gradient method of
Ref. \onlinecite{espig2012}, 
 but we have not tried to use them.

\subsection{Combination of  block power method  with rank reduction: Reduced Rank Block Power Method (RRBPM)  \label{combination}}

Three operations in the  block power method  cause the rank of the basis vectors to increase 
and must therefore be followed by rank reduction.    As already discussed, applying 
 {$\bf{H}$} to a vector increases its rank by a factor of $T$.   Orthogonalization requires adding vectors
and the rank of the sum of two vectors is the sum of their ranks.   
After solving the generalized eigenvalue problem, the basis vectors are updated    by replacing them  with linear 
combinations (the coefficients being elements of the eigenvector matrix) of  basis vectors; this also 
increases the rank.  Orthogonalization and updating  increase the rank by much less than applying 
 {$\bf{H}$}, nevertheless if they are not  followed by a rank reduction the rank of the basis vectors will 
steadily increase during the calculation.

The algorithm we use is:
\begin{enumerate}
\item Define $B$ random rank-one initial vectors ${\bf F}_b$ with elements $\prod_{j=1}^D f^{(1,j)}_{b,i_j}$ for $b=1,\dots,B$, $i_j=1,\dots,n_j$.
\item Orthogonalize the  ${\bf F}_b$     set with a modified Gram-Schmidt procedure adapted to the SOP structure. \label{gram_step}
\item First reduction step: if $B>r$, reduce the rank of  the orthogonalized  
$ {\bf F}_b $ to  $r$ using ALS. \label{first_red}
\item Iterate: 
\begin{enumerate}
\item Apply   ${\bf F}_b         \leftarrow ({\bf H} - \sigma \openone )  {\bf F}_b \; \; \forall b=1,\dots,B$. 
\item Main reduction step: reduce all the   ${\bf F}_b$    to rank $r$ using ALS. 
\item Every $N_{\text{ortho}}$ iterations:
\begin{enumerate}
\item  Orthogonalize the $\{  {\bf F}_b        \}$ set using a  SOP-adapted modified Gram-Schmidt procedure. \label{gram2}
\item Reduce the rank of all the  ${\bf F}_b$        to  $r$ using ALS.\label{red_gram}
\end{enumerate} 
\item Every $N_{\text{diag}}$ iterations (multiple of $N_{\text{ortho}}$): \label{update}
\begin{enumerate}
\item Compute  ${\bf H}^{(\bm{\mathcal{F}})}_{b b'} 
= {\bf F}_b^T   {\bf H}  {\bf  F}_{b'} $ and  ${\bf S}_{b b'} = {\bf F}_b^T {\bf   F}_{b'} $.
\item Solve the generalized eigenvalue problem ${\bf H}^{(\bm{\mathcal{F}})} {\bf U = S U E}$ where  ${\bf E}$ is the diagonal matrix of 
eigenvalues and ${\bf U}$ is the  matrix of  
eigenvectors. \label{diag_step}
\item Update the vectors ${\bf F}^{\text{new}}_{b'}= \sum_{b=1}^B {\bf U}_{bb'} {\bf F}_{b} $.
\item ${\bf F}_b \leftarrow$ reduction of $ {\bf F}^{\text{new}}_{b} $  to rank $r$ using ALS, $\forall b=1,\dots,B$.
\end{enumerate}
\end{enumerate}
\end{enumerate}
At step 2 and step 4(c)i, we orthogonalize with a Gram-Schmidt procedure adapted to exploit the SOP structure 
of the vectors.    This requires computing 
scalar products of 1-D \; $
 {\bf \tilde{f}}^{(1,j)}_{b}$ vectors and adding  $ {\bf F}_{b} $ vectors. At step  4(d)ii  we also add vectors.  Adding vectors
is done by    concatenation.

The program that implements the algorithm is not fully optimized, but we have parallelized some of its steps.
Step 4,a is embarrassingly parallel because $(H - \sigma \openone )$ can be applied to each vector separately. 
In all the  orthogonalization steps, there are two nested loops on  indices $b$ and $b'$:\\

\qquad for $b  =  1$ to $B$ 

\qquad \qquad ${\bf F}_{b} \leftarrow {\bf F}_b / \Vert {\bf F}_b \Vert$

\qquad \qquad for $b' =  b+1$ to $B$ 

\qquad \qquad \qquad { } ${\bf F}_{b'} \leftarrow {\bf F}_{b'} - ({\bf F}_{b}^T  {\bf F}_{b'})  {\bf  F}_{b}$

\qquad \qquad end

\qquad end \\

\noindent The internal loop over    $b'$  is parallelized.   
The reduction of steps 3, 4(b),  4(c) ii, and  4(d)iv is also done in parallel.  

After step \ref{gram_step}, the maximum rank of a basis vector is $B$.
 The first reduction step \ref{first_red} is only done if  $B>r$.
After the application of $(H-\sigma \openone)$, every basis vector has rank
 $Tr$ (or $TB$ during the first passage if $B<r$),  
which is  reduced
 to $r$ in step 4(b).  
After the orthogonalization step \ref{gram2},  as well as after the diagonalization step \ref{diag_step} the 
rank is  $Br$, but is immediately  reduced to $r$.

The memory cost is  the memory required to store the largest rank vectors.
If $T > B$,  then the  largest rank vectors are those obtained after application of 
  $(H-\sigma \openone)$ and they have rank of $Tr$.
If $B > T $,   then the  largest rank vectors are those obtained after 
 the orthogonalization  and they have rank of $Br$.  
Therefore, the memory cost scales as
\bea
\mathcal{O} (n D B T r ) \text{ if } T > B  \label{memory_cost} \\
\mathcal{O} (n D B^2 r) \text{ if } T < B \nonumber ~.
\eea
The CPU  cost is dominated by  the reduction steps 
(scalings are given in section \ref{reduction_section}). 
The cost of  one matrix vector product scales as $\mathcal{O} (T R D n^2)$  (see Eq. \eqref{hpsiproduct}).

\section{Bilinearly coupled harmonic oscillators  \label{models}}

We first test the rank-reduced block power method (RRBPM)  by computing eigenvalues of a simple Hamiltonian
for which exact energy levels are known, 
\begin{equation}
H (q_1,\dots,q_D) =  \sum_{j=1}^D \frac{\omega_j}{2} \left( p_j^2 + q_j^2 \right) + 
\sum_{\substack{i,j=1 \\ i>j}}^D \alpha_{ij} q_i q_j
\label{coupledham}
\end{equation}
with $p_j = - \imath \frac{\partial}{\partial q_j}$.
The product basis $\prod_{j=1}^D \theta_{i_j}^j (q_j)$ is made from
 $n_j$ harmonic oscillator basis functions for each coordinate, i.e.
eigenfunctions of $\frac{\omega_j}{2} \left( p_j^2 + q_j^2 \right)$.
The exact levels of $H$, obtained by transforming to normal coordinates are
\begin{equation}
E_{m_1, m_2, \cdots, m_D} = \sum_{j=1}^D \nu_j \left( \frac{1}{2} + m_j \right)\text{, with } m_j =0,1, \cdots .
\label{theor_spectrum}
\end{equation}
The normal mode frequencies $\nu_j$ are 
 square roots of the eigenvalues of the matrix ${\bf A}$ whose  elements are
 $A_{ii} = \omega_i^2$ and $A_{ij} = \alpha_{ij} \sqrt{\omega_i} \sqrt{\omega_j}$.
Despite the simplicity of the Hamiltonian, it is a good test of the RRBPM.  Our goal is to determine
whether it is possible to obtain accurate levels when the coupling is large enough to significantly 
shift levels.  Is it possible to compute accurate levels with a value of $r$ small enough the memory cost of the 
RRBPM is significantly less than the memory cost of a method that requires storing all $n^D$ components of 
vectors?   Does the ALS procedure for rank reduction work well 
in these conditions?

\subsection{ 6-D coupled oscillators}

We  arbitrarily  choose  the  coefficients  of \Eq{coupledham},
\begin{equation}
\omega_j = \sqrt{j/2}\text{, }j=1,\dots,6.
\end{equation}
For simplicity, the same value $\alpha_{ij}=0.1$ is given to all the coupling constants.  The coefficient of a 
quadratic coupling term is about 14\% 
of the smallest $\omega_j$.  
The  coupling      shifts  the 
 frequencies of transitions $0 \rightarrow 9$ and  
 $0 \rightarrow 39$ by   about 2
percent.  
For realistic transitions around 3000 cm$^{-1}$, this would correspond to a shift of 60 cm$^{-1}$;
 the coupling is therefore significant.   
Energies computed with the parameters of  table~\ref{parametres_temoin} are reported in table ~ \ref{table6D}.
For a given choice of the Hamiltonian parameters and the basis size parameters ($n_j$), 
one expects the accuracy to be limited by 
the values of   $r$,    
$N_{\text{ALS}}$,       
 $B$,         and 
the maximum value of 
 $N_{\text{pow}}$.    
 Increasing any of these will increase the accuracy.  Regardless of the values
of    $r$ and 
$N_{\text{ALS}}$,  accurate energies can be obtained by increasing $B$.   
The    $r$,    
$N_{\text{ALS}}$,       
 $B$,         and 
 Max($N_{\text{pow}}$)  
 values in  table~\ref{parametres_temoin} were determined by testing various values, but 
many sets of parameter values work well.  
  For the bilinearly coupled Hamiltonian,  the largest eigenvalue of 
 {$\bf{H}$} could be estimated from the largest diagonal matrix element, but we compute it with the 
power method (no shift, block size of one).    We choose a shift close to  the largest eigenvalue. 
This shift  is not the optimal value given in Eq. \eqref{idealshift}.
Decreasing the shift, as explained in section \ref{method},
 slightly accelerates the convergence.
For  3000  iterations the calculation  requires   approximately 6 min on a computer with 
2 Quad-Core AMD  2.7 GHz processors, using all of the processing cores.

%
\begin{table}[ht]
\caption{
Parameters for the 
 calculation with $D=6$.}
\begin{tabular}{ll}
\hline
\hline
$D$           &6 \\
$\omega_j$   &  $\sqrt{j/2}$ \\
$\alpha_{ij}$  & 0.1 \\
$n_j,\quad \forall j=1,\dots,6$         & 10 \\
Reduction rank $r$      & 10 \\
$N_{\text{ALS}}$        & 10 \\
Block size $B$          & 40 \\
Maximum $N_{\text{pow}}$     & 3000 \\
$E_{\text{max}}$ estimate  & 80.36 \\
Energy shift $\sigma$    & 81  \\
$N_{\text{ortho}}$        &  20 \\
$N_{\text{diag}}$        &  20 \\
\hline
\hline
\end{tabular}
\label{parametres_temoin}
\end{table}

\begin{table*}[ht]
\caption{Energy levels 
 of the 6D coupled oscillator Hamiltonian.  
 From left to right: energy label $n$, exact energy, RRBPM energy with no updating,  relative error,
 RRBPM energy with  updating,  relative error.
The last column is the normal mode assignment. Parameters from table \ref{parametres_temoin} are used.}
\begin{tabular}{ccccccccc}
\hline
\hline  
$n$ & $E_{n,\text{th}}$ & $E_{n,\text{num}}^{(1)}$ (no update) & $\frac{E_{n,\text{num}}^{(1)}-E_{n,\text{th}}}{E_{n,\text{th}}}$ & $E_{n,\text{num}}^{(2)}$ (with update) & $\frac{E_{n,\text{num}}^{(2)}-E_{n,\text{th}}}{E_{n,\text{th}}}$  & Assignment \rule[-8pt]{0pt}{22pt}\\
\hline
 0 & 3.8164041 & 3.8164063 & 6.0$ \times 10^{-7}$ & 3.8164053 & 3.1$ \times 10^{-7}$ & - \\
 1 & 4.5039223 & 4.5039260 & 8.2$ \times 10^{-7}$ & 4.5039269 & 1.0$ \times 10^{-6}$ & $\nu_1$ \\
 2 & 4.7989006 & 4.7989650 & 1.3$ \times 10^{-5}$ & 4.7989351 & 7.2$ \times 10^{-6}$ & $\nu_2$ \\
 3 & 5.0262787 & 5.0262979 & 3.8$ \times 10^{-6}$ & 5.0263184 & 7.9$ \times 10^{-6}$ & $\nu_3$ \\
 4 & 5.1914405 & 5.1914748 & 6.6$ \times 10^{-6}$ & 5.1914768 & 7.0$ \times 10^{-6}$ & $2 \, \nu_1$ \\
 5 & 5.2196500 & 5.2196889 & 7.4$ \times 10^{-6}$ & 5.2197161 & 1.3$ \times 10^{-5}$ & $\nu_4$ \\
 6 & 5.3938508 & 5.3940540 & 3.8$ \times 10^{-5}$ & 5.3939608 & 2.0$ \times 10^{-5}$ & $\nu_5$ \\
 7 & 5.4864188 & 5.4865101 & 1.7$ \times 10^{-5}$ & 5.4864995 & 1.5$ \times 10^{-5}$ & $\nu_1+\nu_2$ \\
 8 & 5.5886301 & 5.5886564 & 4.7$ \times 10^{-6}$ & 5.5887086 & 1.4$ \times 10^{-5}$ & $\nu_6$ \\
 9 & 5.7137969 & 5.7139860 & 3.3$ \times 10^{-5}$ & 5.7141074 & 5.4$ \times 10^{-5}$ & $\nu_2+\nu_3$ \\
 \vdots & \vdots & \vdots &\vdots &\vdots &\vdots &\vdots \\ 
 19 & 6.3763473 & 6.3797375 & 5.3$ \times 10^{-4}$ & 6.3770751 & 1.1$ \times 10^{-4}$ & $\nu_2+\nu_5$ \\
 20 & 6.4013151 & 6.4041135 & 4.4$ \times 10^{-4}$ & 6.4017330 & 6.5$ \times 10^{-5}$ & $2\,\nu_1+\nu_3$ \\
 21 & 6.4295246 & 6.4331587 & 5.7$ \times 10^{-4}$ & 6.4305409 & 1.6$ \times 10^{-4}$ & $\nu_3+\nu_4$ \\
 22 & 6.4689153 & 6.4728294 & 6.1$ \times 10^{-4}$ & 6.4693700 & 7.0$ \times 10^{-5}$ & $\nu_1+2\,\nu_2$ \\
 23 & 6.5664770 & 6.5686726 & 3.3$ \times 10^{-4}$ & 6.5665373 & 9.2$ \times 10^{-6}$ & $4\,\nu_1$ \\
 24 & 6.5711267 & 6.5731186 & 3.0$ \times 10^{-4}$ & 6.5720287 & 1.4$ \times 10^{-4}$ & $\nu_2+\nu_6$ \\
 \vdots &\vdots &\vdots &\vdots &\vdots &\vdots &\vdots \\ 
 34 & 6.8896648 & 6.9025957 & 1.9$ \times 10^{-3}$ & 6.8924037 & 4.0$ \times 10^{-4}$ & $\nu_1+\nu_2+\nu_4$ \\
 35 & 6.9236715 & 6.9354401 & 1.7$ \times 10^{-3}$ & 6.9253179 & 2.4$ \times 10^{-4}$ & $\nu_1+2\,\nu_3$ \\
 36 & 6.9636666 & 6.9665546 & 4.1$ \times 10^{-4}$ & 6.9645821 & 1.3$ \times 10^{-4}$ & $2\,\nu_1+\nu_6$ \\
 37 & 6.9712976 & 6.9796692 & 1.2$ \times 10^{-3}$ & 6.9721608 & 1.2$ \times 10^{-4}$ & $2\,\nu_5$\\
 38 & 6.9912717 & 6.9990297 & 1.1$ \times 10^{-3}$ & 6.9928003 & 2.2$ \times 10^{-4}$ & $2\,\nu_2+\nu_3$ \\
 39 & 6.9918761 & 7.0057516 & 2.0$ \times 10^{-3}$ & 6.9947683 & 4.1$ \times 10^{-4}$ & $\nu_4+\nu_6$ \\
\hline
\hline
\end{tabular}
\label{table6D}
\end{table*}

\begin{figure}[htp]
\includegraphics[width=\linewidth]{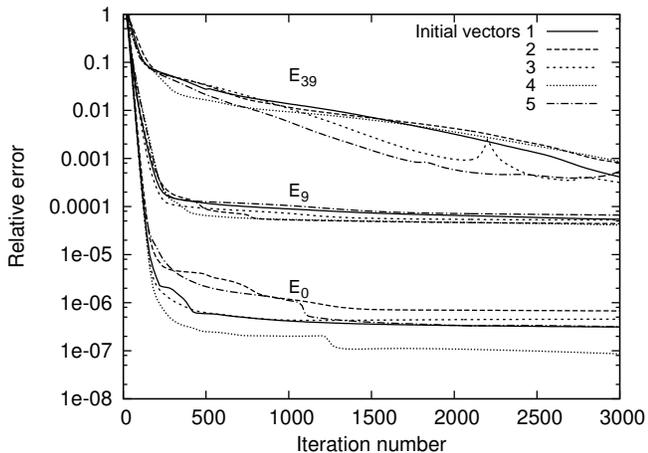}
 \caption{Convergence curves for the lowest eigenvalue $E_0$, the highest eigenvalue of the block $E_{39}$ and another one inside the block, $E_9$.  The y axis is the relative error (logarithmic scale)  and
  the x axis is the iteration number, $N_{\text{pow}}$. Results are shown for five    calculations 
 using five different randomly chosen sets of start vectors.}  
\label{fig01_6d}
\end{figure}

\begin{figure}[htp]
\includegraphics[width=\linewidth]{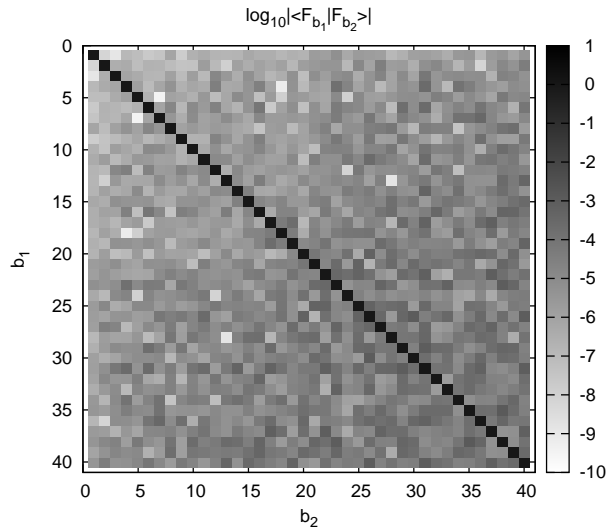}
 \caption{Grey scale logarithmic representation of the overlap matrix, $\log_{10} \vert {\bf S}_{b b'} \vert = \log_{10} \vert {\bf F}_b^T {\bf F}_{b'} \vert $ after the last (orthogonalization + reduction) step. The black pixels correspond to approximately~1.}
\label{fig02_6d}
\end{figure}

\begin{figure*}
\subfigure[][]{%
\label{fig03a_6d_leclerc}%
\includegraphics[width=0.5\linewidth]{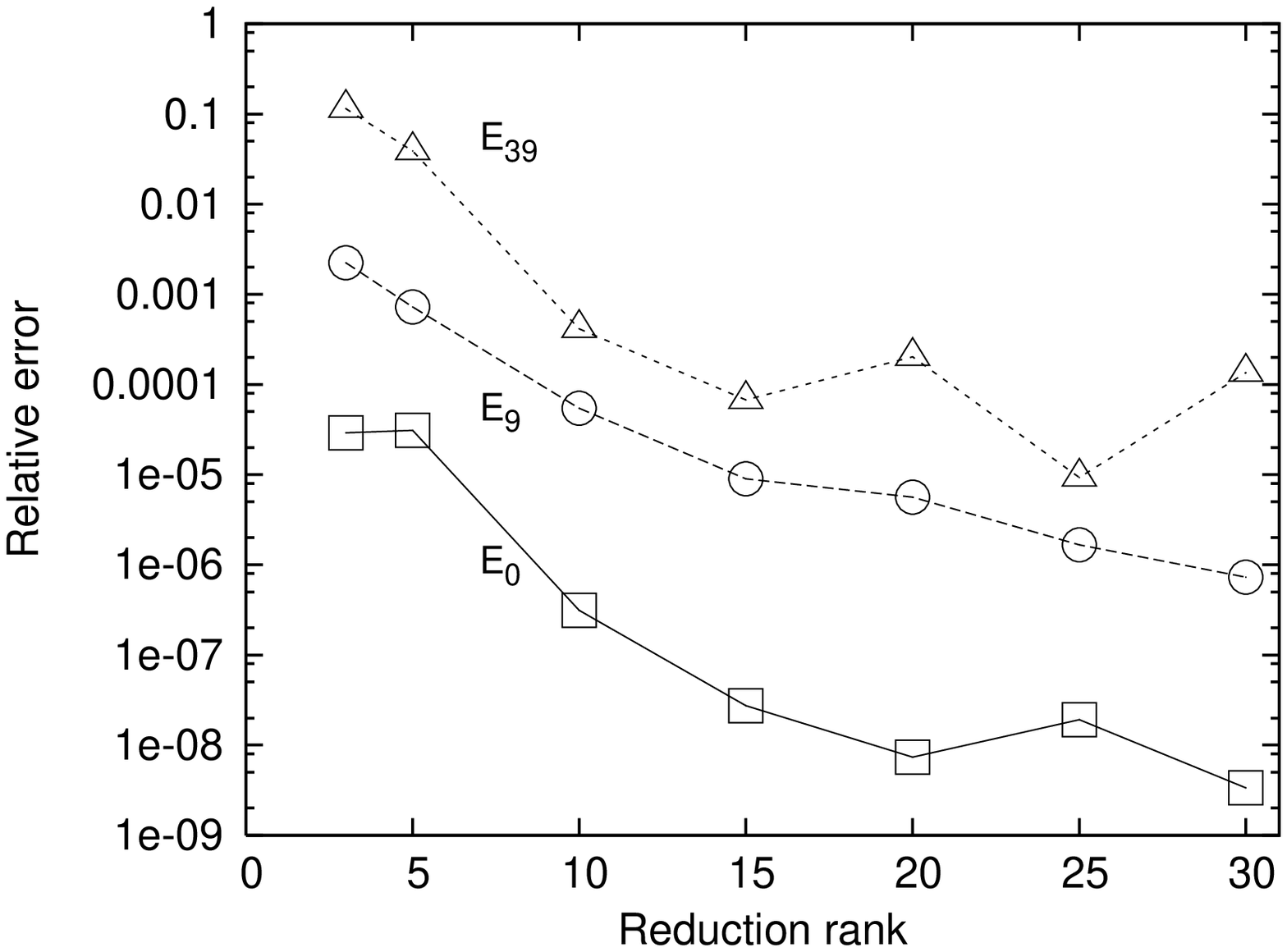}}%
\subfigure[][]{%
\label{fig03b_6d_leclerc}%
\includegraphics[width=0.5\linewidth]{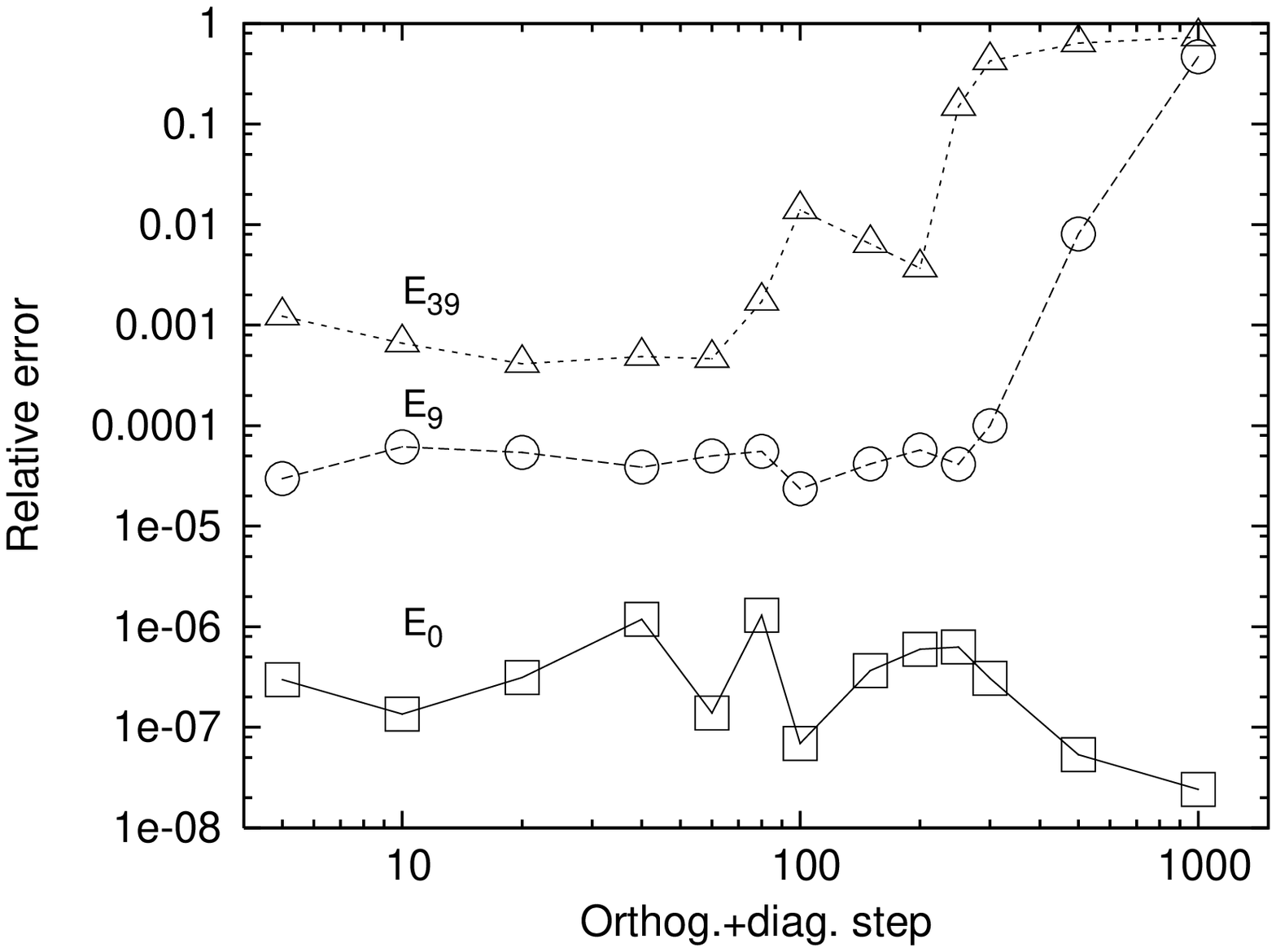}}\\
\subfigure[]{%
\label{fig03d_6d_leclerc}%
\includegraphics[width=0.5\linewidth]{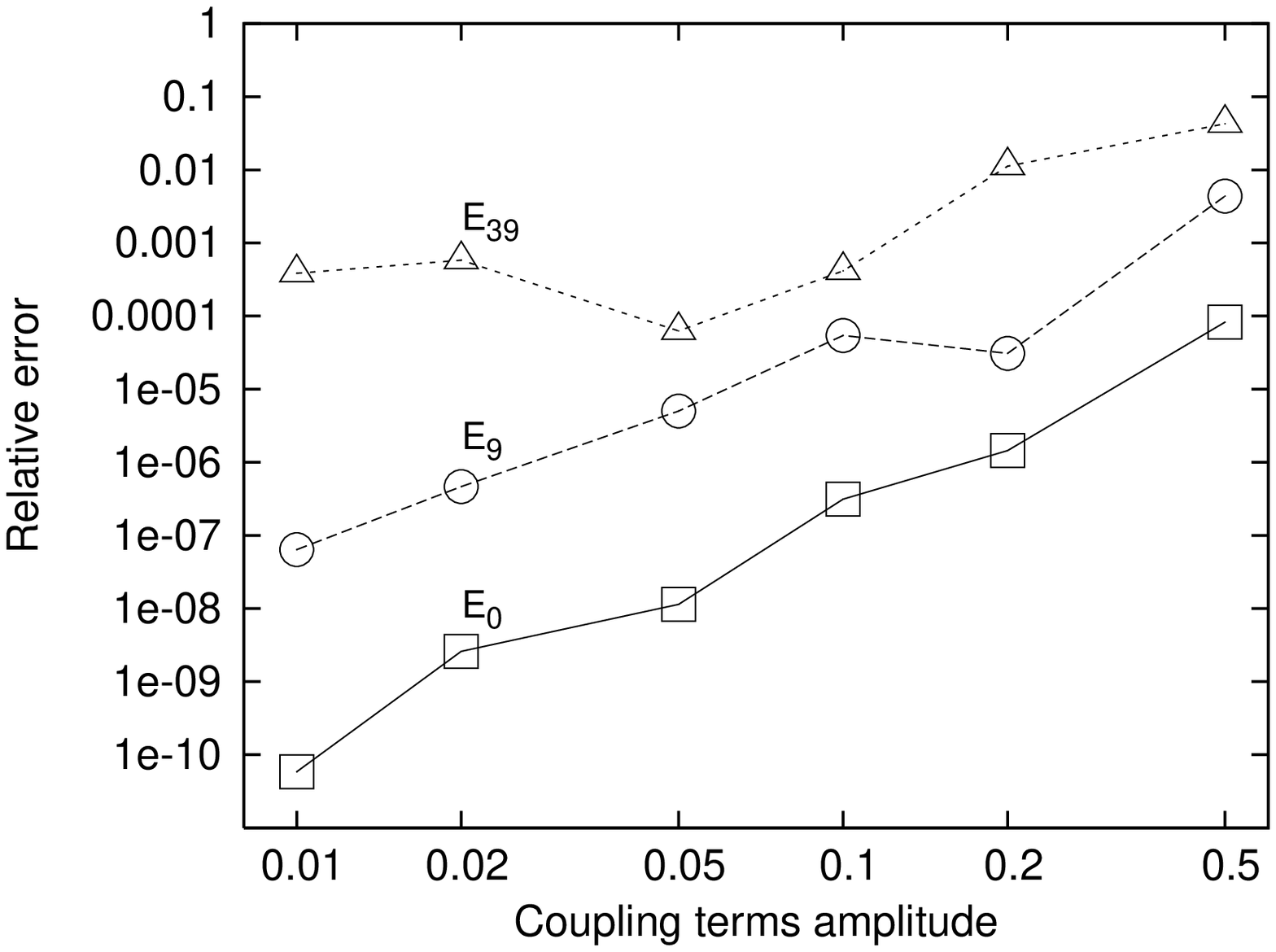}}
\caption[]{%
 Relative error of $E_0$, $E_9$ and $E_{39}$ after Max($N_{\text{pow}}$)=3000 RRBPM iterations, on
 a logarithmic scale and as a function of: (a) the reduction rank $r$; (b) the orthogonalization-diagonalization step
size  (here $N_{\text{ortho}} = N_{\text{diag}}$); 
(c) the coupling term amplitude $\alpha_{ij}$.}%
\label{fig03_6d}%
\end{figure*}

Two versions of the algorithm are tested, with and  without  updating   eigenvectors (item \ref{update} 
in the algorithm of section \ref{combination}). 
When there is no updating during the iterations, the subspace diagonalization is done from time to time, only to follow the convergence of the results. 
With the parameters of  table \ref{table6D} and implementing updating,  
  the zero-point energy (ZPE) $E_0$ is recovered with a $10^{-7}$ 
relative error (6 significant digits). 
For low-lying levels there   
 is no significant difference between energies obtained with and without updating.
As expected, the lowest eigenvalues are the most accurately determined,  with a relative error  of less than $10^{-4}$ 
for the first ten eigenvalues. 
For the higher eigenvalues  in 
 the block, a little less than one order of magnitude is gained by updating.
The quality of the highest eigenvalues of the block is less good,  but even the largest  calculated eigenvalue $E_{39}$
has a relative error of order $10^{-4}$, despite the very low rank that we have chosen ($r=10$) and its  proximity
 with neighboring eigenvalues. 
These results confirm that even with basis functions with only $r=10$ terms,  good accuracy can be achieved.   
With these parameters, the calculation requires only 20 MB of memory. 

To verify that accuracy is not limited by  $N_{\text{pow}}$ we plotted, in   Fig. \ref{fig01_6d},
 the relative error for three eigenvalues as a function of   $N_{\text{pow}}$,  
 with five different start  vectors.  
We focus  attention on the ZPE,  $E_0$, the highest state of the block, $E_{39}$ and an intermediate 
eigenvalue, $E_9$.
As expected, the lowest eigenvalue converges faster than $E_9$ and $E_{39}$.
Error decreases rapidly for  the first 250 iterations
and thereafter more slowly. The results depend little on the choice of the start
 vector  (maximum variation of about an  order of 
magnitude after 3000 iterations).

%
As explained in the last section, it is crucially important to reduce the rank of the SOP vectors.   However, the reduction
changes the vectors  and the space they span.  
To assess how much the functions are changed by reduction,  we examine the   overlap matrix 
 {$\bf{S}$}  after the reduction following 
orthogonalization. 
The overlap matrix after
 iteration number 3000  is shown in Fig. \ref{fig02_6d}.  The  reduction reduces the rank from $Br$ to $r$. 
Reduction causes this 
matrix to differ from an identity matrix.  
Each pixel corresponds  to 
one element of the overlap matrix 
$S_{b b'}$ with a logarithmic grey-scale. After normalizing the diagonal elements to one,  
 the largest  off-diagonal elements        are
 of order $10^{-3}$.
Neglecting the off-diagonal elements
 creates sufficient errors in the spectrum to justify the choice of a generalized eigenvalue algorithm in the algorithm described in section \ref{combination}.

Energy errors in this section are differences   between  RRBPM energies  and exact (from normal frequencies) energies.
In order to test the ability of the RRBPM to determine eigenvalues of a basis representation of the Hamiltonian operator,
it would be better to take differences of  RRBPM energies and exact eigenvalues of 
${\bf  H } $.   By computing exact eigenvalues of ${\bf  H } $ (using a product basis Lanczos method),
we have verified that the accuracy of RRMP energies in not 
 limited by the primitive basis.  
The Lanczos eigenvalues agree with the exact energies to within    $10^{-10}$.   The Lanczos calculation is easy in 
6D.

To understand how changing the parameters in table   
 \ref{parametres_temoin}
 affects energy levels, we did a series of 
calculations keeping all but one of the parameters fixed at the values in table 
 \ref{parametres_temoin}. 
 These tests are done
for a single start vector.
%
In Fig. \ref{fig03a_6d_leclerc} we show how the relative error, for three energies,  decreases as $r$ is increased.  
Energies are, of course,     more accurate when the reduction rank is larger.
However, the error vs $r$ curves flatten as $r$ increases.  Even with relatively small values of $r$, good accuracy 
is obtained, e.g., the relative error of the ZPE is  $3\times 10^{-5}$  
with $r=5$.  
The most costly part of the calculation is the reduction  
 which involves a  solution of linear equations scaling as $r^3$.  
 Fig. \ref{fig03b_6d_leclerc} shows that although  the error increases as $N_{\text{diag}}=N_{\text{ortho}}$ is increased,
   reorthogonalizing and updating  less frequently  does not significantly degrade the accuracy.  
Starting at about $N_{\text{diag}} > 100$, the error does increase somewhat.
The ZPE is accurate  even with a  large $N_{\text{diag}}$ (without any orthogonalization, all the basis 
 vectors would tend to the lowest eigenstate).
It is fortunate that $N_{\text{diag}}=N_{\text{ortho}}$ need not be large because except for  updating, 
everything can be parallelized over the vectors in the block. 
The subspace diagonalization requires  $\mathcal{O}(B^2)$ scalar products to compute the matrices 
and $\mathcal{O}(B^3)$ operations for the direct diagonalization. 
However these operations are not costly  because they are repeatedly infrequently.  
Another important parameter is the number of ALS iterations 
(see section \ref{reduction_section}).   
We have used ALS to reduce the rank of test
functions and observed that the error vs number of ALS iterations curve often flattens and then 
decreases slowly 
(not shown).  
The convergence behavior also depends on the vector whose rank is being reduced.  It is 
therefore extremely likely that 
some of the reduced rank vectors we use as basis vectors in  the RRBPM, where we fix  the number of ALS iterations,
have significant errors.  The fact that the basis vectors differ from those one would have with a 
standard block power method is not necessarily a cause for concern because we project into the space 
spanned by the basis vectors.   This works for the same reason that it is possible to use inexact spectral 
transforms.\cite{huang2000,poirier2001,poirier2002,billpist,jc} 
We have nevertheless done tests with several values of $N_{ALS}$.  With  $N_{ALS}$= 10
accurate energies are obtained.  
If  $N_{ALS}$ is too small, errors are larger and the errors of higher 
states are larger.  None of the 39 levels we compute are significantly improved by doubling  $N_{ALS}$.
Replacing  $\sigma=81$ with  $\sigma=44$, 
consistent with the optimal value of Eq.~\eqref{idealshift}  slightly accelerates  the convergence.
Increasing the block size $B$ accelerates   convergence, but the effect is most important when the 
iteration number is less than about 500.   

%

Of course as the coupling    $\alpha_{ij}$ is increased the accuracy of a method based on rank reduction
must degrade.   
The    $\alpha_{ij}$ value we have used  forces a realistic mixing.  Although low-rank approximations for the 
lowest states are good, many of the  $\prod_{j=1}^D \theta_{i_j}^j (q_j)$ basis functions are mixed. 
Nevertheless, to force the RRBPM to fail, as it must, we increased  $\alpha_{ij}$.   The results are 
displayed in Fig. \ref{fig03d_6d_leclerc}.  All other parameters have the values in 
table \ref{parametres_temoin}
except the energy shift which has been adapted for each run because the spectral range changes. 
As expected, increasing  the coupling increases the error. 
 When  $\alpha_{ij}$ is larger than 0.1  better accuracy can be obtained by increasing $r$. 

\subsection{%
 20-D coupled oscillators}

The basis used in  the  6D calculation of the previous subsection has $10^6$ functions.  Obviously,
direct diagonalisation methods cannot be used,   but it is easy compute eigenvalues of 
 {$\bf{H}$}   with the Lanczos algorithm.    In this subsection, we present results for a 20D calculation with 
the Hamiltonian of \Eq{coupledham}  and the parameters listed in table \ref{parametres_20d}.
  In 
this case, the Lanczos calculation is also impossible.  With $n_j=10$, a  single vector has $10^{20}$ components and to 
keep a vector in memory one would need $ \sim 8 \times 10^{11}$ GB.  Using the RRBPM and less than 1 GB, we compute 
about 50 
energy levels with relative accuracy of about $10^{-5}$.    The Hamiltonian has 210 terms. 
The shift value is slightly larger than half  the maximum eigenvalue ($\simeq 568$).
The shift value is an  optimal shift (Eq. \eqref{idealshift}), obtained from  
  $E_{\text{max}}$ and    $E_>$.  
To determine  $E_>$, a few iterations of the RRBPM are done using the non-optimal shift $\sigma \simeq E_{\text{max}}$ and a block of size $B+1$.
\begin{table}[ht]
\caption{Numerical parameters for the calculations with $D=20$ coupled oscillators.}
\begin{tabular}{ll}
\hline
\hline
$D$          & 20 \\
$\omega_j$   &  $\sqrt{j/2}$ \\
$\alpha_{ij}$  & 0.1 \\
$n_j, \quad \forall j=1,\dots,20$  & 10 \\
Reduction rank $r$      & 20 \\
$N_{\text{ALS}}$        & 10 \\
Block size $B$          & 56 \\
Maximum $N_{\text{pow}}$& 5000 \\
$E_{\text{max}}$ estimate  & 568 \\
Energy shift $\sigma$    &  320 \\
$N_{\text{ortho}}$        & 40 \\
$N_{\text{diag}}$        &  40 \\
\hline
\hline
\end{tabular}
\label{parametres_20d}
\end{table}

\begin{table}[ht]
\caption{Energy levels  of the  20 coupled oscillator Hamiltonian of Eq.~\eqref{coupledham}.  From left to right:  energy level  number, exact energy level,  RRBPM energy level, relative error,  normal mode
assignment.}
\begin{tabular}{ccccc}
\hline
\hline
$n$ & $E_{n,\text{th}}$ & $E_{n,\text{num}}$ & $\frac{E_{n,\text{num}}-E_{n,\text{th}}}{E_{n,\text{th}}}$ & Assignment  \rule[-8pt]{0pt}{22pt} \\
\hline
0 & 21.719578 & 21.719587 & 4.2$\times 10^{-7}$ & - \strut\\
1 & 22.398270 & 22.398294 & 1.1$\times 10^{-6}$ & $\nu_1$\\
2 & 22.691775 & 22.691826 & 2.2$\times 10^{-6}$ & $\nu_2$\\
3 & 22.917012 & 22.917129 & 5.1$\times 10^{-6}$ & $\nu_3$\\
4 & 23.076962 & 23.077014 & 2.3$\times 10^{-6}$ & $2\,\nu_1$\\
5 & 23.106960 & 23.107006 & 2.0$\times 10^{-6}$ & $\nu_4$\\
6 & 23.274380 & 23.274502 & 5.3$\times 10^{-6}$ & $\nu_5$\\
7 & 23.370467 & 23.370629 & 6.9$\times 10^{-6}$ & $\nu_1 + \nu_2$\\
8 & 23.425814 & 23.425951 & 5.8$\times 10^{-6}$ & $\nu_6$\\
9 & 23.565153 & 23.565222 & 2.9$\times 10^{-6}$ & $\nu_7$\\
\vdots &\vdots &\vdots &\vdots &\vdots \\
20 & 24.049160 & 24.049374 & 8.9$\times 10^{-6}$ & $2\,\nu_1+\nu_2$\\
21 & 24.079158 & 24.079914 & 3.1$\times 10^{-5}$ & $\nu_3+\nu_4$\\
22 & 24.104506 & 24.104878 & 1.6$\times 10^{-5}$ & $\nu_1 + \nu_6$\\
23 & 24.114446 & 24.114570 & 5.2$\times 10^{-6}$ & $2\,\nu_3$\\
\vdots &\vdots &\vdots &\vdots &\vdots \\
30 & 24.342665 & 24.343080 & 1.7$\times 10^{-5}$ & $\nu_1+2\,\nu_2$\\
31 & 24.346217 & 24.346365 & 6.1$\times 10^{-6}$ & $\nu_{14}$\\
32 & 24.373625 & 24.373996 & 1.5$\times 10^{-5}$ & $\nu_1+\nu_8$\\
33 & 24.398012 & 24.398676 & 2.7$\times 10^{-5}$ & $\nu_2+\nu_6$\\
\vdots &\vdots &\vdots &\vdots &\vdots  \\
40 & 24.532333 & 24.533376 & 4.3$\times 10^{-5}$ & $\nu_{16}$\\
41 & 24.537351 & 24.539130 & 7.3$\times 10^{-5}$ & $\nu_2+\nu_7$\\
42 & 24.567902 & 24.570246 & 9.5$\times 10^{-5}$ & $\nu_1 +\nu_2 +\nu_3$\\
43 & 24.611100 & 24.613013 & 7.8$\times 10^{-5}$ & $\nu_1+\nu_{10}$\\
\vdots &\vdots &\vdots &\vdots &\vdots \\
50 & 24.709939 & 24.725314 & 6.2$\times 10^{-4}$ & $\nu_{18}$\\
51 & 24.721068 & 24.765667 & 1.8$\times 10^{-3}$ & $\nu_1 + \nu_{11}$\\
52 & 24.727852 & 24.786084 & 2.4$\times 10^{-3}$ & $3\,\nu_1 + \nu_2$\\
53 & 24.757850 & 24.810515 & 2.1$\times 10^{-3}$ & $\nu_1 + \nu_2 + \nu_4$\\
54 & 24.762587 & 24.823982 & 2.5$\times 10^{-3}$ & $\nu_3 + \nu_7$\\
55 & 24.783198 & 24.863299 & 3.2$\times 10^{-3}$ & $2\,\nu_1 + \nu_6$\\
\hline
\hline
\end{tabular}
\label{table20D}
\end{table}

\begin{figure}[htp]
\includegraphics[width=\linewidth]{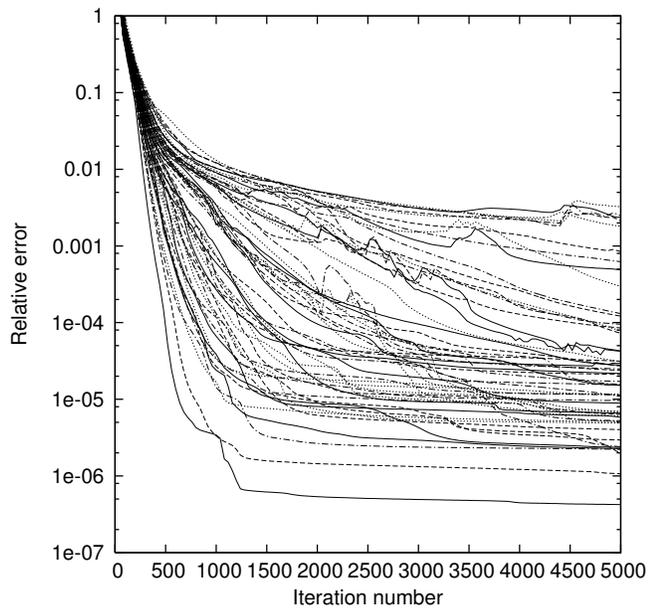}
 \caption{Convergence curves of  the energy levels   
 of the 20-D  coupled oscillator Hamiltonian.  
 Parameters of table \ref{parametres_20d} are used.  The y-axis is
 the relative error on a logarithmic scale and the x-axis is the iteration number $N_{\text{pow}}$.}
\label{fig04_20d}
\end{figure}

%

Some computed eigenvalues are  presented in table \ref{table20D} along with the corresponding 
exact eigenvalues calculated  from the normal mode frequencies.  Energies are assigned with normal mode
frequencies.  
 In this case it is not possible to compare the RRBPM results with those obtained from a standard 
Lanczos calculation due to the memory cost of the latter.   We therefore have no choice but to calculate  
relative errors  using  exact eigenvalues, 
assuming that the finite basis introduces no significant error, as shown in the 6D case. 
The agreement between theoretical and numerical eigenvalues is good,  with errors of the    order  of 
$10^{-5}$,  and especially
 good for the lowest energies.

Convergence curves are  presented in Fig. \ref{fig04_20d}, which shows the relative error for all  56 eigenvalues 
of a $B \times B$ block as a function of the iteration number.   
 The error does not  decrease monotonically because rank reduction can worsen the basis. The smallest energies are 
most accurate.  The largest energies are not converged with respect to the number of iterations even with 
a maximum value of $N_{\text{pow}}$ equal to  5000.
Most of the eigenvalues stabilize with an error less than   about $10^{-4}$.  %
Although this figure is similar to  Fig. \ref{fig01_6d}, convergence is slower. 
Increasing the number of
coordinates from 6 to 20 increases the spectral range of  {$\bf{H}$} and slows the convergence of the block 
power method.  It also makes wavefunctions more complex and reduces the accuracy of low-rank vectors.  
Ideally, the cost of  the calculation  would scale linearly with $D$. 
 In practice, linear scaling 
 is not realized because increasing the 
 number of coupled degrees of freedom increases the rank required to represent wavefunctions.
This 
calculation, with 5000 iterations 
takes 2.5 days using 14 processors.   
The memory  cost  is a factor of 50 (consistent with  \Eq{memory_cost})  
  greater than in the 6D case  because the 20D Hamiltonian has many more terms than the 6D Hamiltonian, but is still  
   less than 1 GB. 

\section{Application to a 12-D Model for acetonitrile  (CH$_3$CN) \label{ch3cn}}

It is extremely encouraging that with the RRBPM it is possible to calculate accurate 
energies of a 20D Hamiltonian.   The Hamiltonian of the previous section is chosen to 
facilitate testing the numerical approach.  Does the RRBPM also work well for a realistic
Hamiltonian?   One might worry that rank reduction will make it impossible to compute accurate energies
for a Hamiltonian with a large number of coupling terms with realistic magnitudes.  
 In this section we confirm that it works well when applied to a 12D  Hamiltonian in 
normal coordinates with  a quartic potential.  The potential is for 
      acetonitrile  CH$_3$CN.

The normal coordinates are labelled  $q_k$ with $k=1, 2, \cdots 12$.    $q_k, ~ k=5,6, \cdots $ are  two-fold degenerate
 (they correspond to $q_{5}\dots q_{8}$ of Ref. \onlinecite{begue2005}).  We use the $J=0$ normal coordinate 
 kinetic energy operator (KEO), but omit the $\pi-\pi$ cross terms and the potential-like term, 
\cite{watson1968}
\begin{equation}
K = - \frac{1}{2} \sum_{k} \omega_k \frac{\partial ^2}{\partial q_k^2}. 
\label{watsonhamiltonian}
\end{equation}

We use the  quartic potential of 
 Ref. \onlinecite{avila2011} which is inferred  from Ref. \onlinecite{begue2005}.
Ref. \onlinecite{begue2005} gives  force constants   calculated 
with a hybrid coupled cluster/density functional theory method.  They also report 
 vibrational levels  computed  from  subspaces determined with  second  order perturbation theory.
The potential is  
\begin{eqnarray}
V (q_1,\dots,q_{12})&=& 
\frac{1}{2} \sum_{i=1}^{12} \omega_i q_i^2
+\frac{1}{6} \sum_{i=1}^{12}\sum_{j=1}^{12} \sum_{k=1}^{12} \phi^{(3)}_{ijk} q_i q_j q_k \nonumber \\
&&+\frac{1}{24} \sum_{i=1}^{12} \sum_{j=1}^{12} \sum_{k=1}^{12} \sum_{\ell=1}^{12} \phi^{(4)}_{ijk\ell} q_i q_j q_k q_{\ell}~.
\label{pot}
\end{eqnarray}
According to  Ref. \onlinecite{begue2005}, constants smaller than 
 6 cm$^{-1}$ were not reported.  
%
The force constants must satisfy symmetry relations given by Henry and Amat \cite{henry1960,henry1965},
but usually expressed in terms of the Nielsen $k$ constants\cite{nielsen1951}
that correspond to the $\phi$ in \Eq{pot}. 
   All of the (937) non-zero $\phi$ constants can be determined from (358)
 constants that Henry and Amat denote $k^0$, $k^1$,  and $k^2$.  
 Ref. \onlinecite{begue2005}  reports    132   $\phi$.   Some of the missing  $\phi$ 
are less than  6 cm$^{-1}$, some of the missing   $\phi$   are not small and can be determined from those 
reported, some of     the missing  $\phi$   are not small and cannot be determined
from those reported.
Avila et al.  assume that the force constants reported in  Ref. \onlinecite{begue2005} are 
the  $\phi$ that correspond to 
the $k^0$ force constants
(the corresponding $\phi$ are derivatives of the potential 
 with respect to the $x$ components of the doubly degenerate 
normal coordinates)
and, because they do not have values for them, put  the $k^1$   and $k^2$ force constants equal to zero. 
The resulting  potential is invariant with respect to  
the  $C_{3v}$ operations.  
There are   299 (108 cubic  and 191 quartic)   coupling terms  in the potential.
 Poirier and Halverson  infer a different potential from the force constants published in 
 Ref. \onlinecite{begue2005}.\cite{privatecomm}  
 Most energy levels on their potential differ from 
their counterparts   on the potential of  Ref. \onlinecite{avila2011} by less than 1 cm$^{-1}$.  Either potential
could be used to test the RRBPM, but we choose the potential of   Ref. \onlinecite{avila2011}.

To make the    
 $\prod_{j=1}^D \theta_{i_j}^j (q_j)$ basis  we use 
 the harmonic oscillator basis functions that are eigenfunctions of the quadratic part of the 
Hamiltonian.  The harmonic frequencies are  \cite{begue2005} 
(in cm$^{-1}$) $\omega_1=3065$, $\omega_2=2297$, $\omega_3=1413$, $\omega_4=920$,  $\omega_5=\omega_6=3149$, $\omega_7=\omega_8=1487$, $\omega_9=\omega_{10}=1061$, 
$\omega_{11}=\omega_{12}=361$.

The RRBPM  is used with the parameters of  table \ref{parametres_ch3cn} to compute the smallest 70 
energies of CH$_3$CN.   More energy levels could be obtained by using a larger block ($B$) or 
by combining the  RRBPM  with the preconditioned inexact spectral transform method.
\cite{huang2000,poirier2001}
 Increasing $B$ would 
obviously increase the memory cost of the calculation,  but with $B=70$ the memory cost is less than 
1 GB. 
In the direct product basis, storing only one vector would take 1113     GB. 
We have not incorporated symmetry adaptation and degenerate levels are calculated together.
We use the same  $n_j$ as in    Ref. \onlinecite{avila2011}
(but no pruning).  The  $n_j$   take  into account the harmonic frequencies 
 $\omega_j$ and some strong  coupling terms.

\begin{table}[ht]
\caption{%
Parameters for the CH$_3$CN calculations.}
\begin{tabular}{ll}
\hline
\hline
$D$          & 12 \\
$n_j, j=\{1,3,4,5,6,9,10\}$  & 9 \\
$n_j, j=\{2,7,8\}$  & 7 \\
$n_j, j=\{11,12\}$  & 27 \\
Reduction rank $r$      & 20 \\
$N_{\text{ALS}}$        & 10 \\
Block size $B$          & 70 \\
Maximum $N_{\text{pow}}$& 6000 \\
$E_{\text{max}}$ estimate  & 311419 \\
Energy shift $\sigma$    &  170000 \\
$N_{\text{diag}}$        & 20 \\
\hline
\hline
\end{tabular}
\label{parametres_ch3cn}
\end{table}

The lowest energy levels are 
 given in table \ref{resultsCH3CN} where they  are compared to previous theoretical 
results.    
We have  determined 
vibrational assignments  in two  ways. Note that in the previous section each level was assigned the 
label of an exact wavefunction but that in this section we assign, in the more usual sense, each level to 
the $\prod_{j=1}^D \theta_{i_j}^j (q_j)$ basis function that most nearly approximates it.  
First, we use rank reduction to assign.  
Columns of  {$\boldsymbol{\mathcal{F}}$}  {$\bf{U}$} =${\boldsymbol{ \mathcal{F}}}^{\text{new}}$
which approximate columns of  {$\bf{V}$},   where  {$\bf{H} \bf{V}$}   =   {$\bf{V} \bf{E}$},   
 have rank $Br$.  Using the ALS algorithm each  is reduced to a rank one vector  $ \prod_{j=1}^D f^{(1,j)}_{i_j} $ 
that 
approximates
a column of  {$\bf{V}$}.  A level is assigned to $v_j, v_{j'}, v_{j''}, \cdots$ if the factors
 $f^{(1,j)}_{v_j}$  $f^{(1,j')}_{v_{j'}}$,  $f^{(1,j'')}_{v_{j''}}$ etc are large and all other factors are small.     
Second, we assign on the basis of components of 
columns of  ${\boldsymbol{ \mathcal{F}}}^{\text{new}} = {\boldsymbol{{\mathcal{F}}}} {\bf{U}}$.
A column of  ${\boldsymbol{ \mathcal{F}}}^{\text{new}} $ is a linear combination of vectors of the form  
of   Eq. \eqref{sop} and hence also of the form of   Eq. \eqref{sop}.   Using the RRBPM it 
is possible to compute energy levels without ever actually evaluating the sum in  Eq. \eqref{sop}, however, 
to assign we need elements of    ${\boldsymbol{\mathcal{ F}}}^{\text{new}} $ and must therefore do the sum.   
To identify dominant components, we  
 calculate only the elements $F^{\text{new}}_{b, i_1\dots i_D}$ suspected to be large.  They are those 
for which  $\sum_{j=1}^{n_j} i_j$ is small.  
When the two assignment   procedures yield different results  and when no assignment can be deduced from 
the first we use the second.

Owing to the fact that we use the same basis and same potential as  Ref. \onlinecite{avila2011}, 
differences  between our energies and those of  Ref. \onlinecite{avila2011} are a measure of the accuracy of the RRBPM.  
The energies of 
  Ref. \onlinecite{avila2011}  are obtained using a method that  requires
about an order of magnitude 
 more memory, but which does not require 
a simple force-field type potential.  Experimental values are also listed  in the last column.  The RRBPM energies 
are close to those of    Ref. \onlinecite{avila2011}.  The lowest RRBPM levels in 
 table \ref{resultsCH3CN} differ from the numbers in    Ref. \onlinecite{avila2011} by a few cm$^{-1}$.  The higher
levels differ more.   The levels with the largest errors are:  
 1779.88, 1780.66 and 2000.43, 2007.90 and the seven 
highest eigenvalues of the block.  The highest eigenvalues could be improved by using a larger $B$.  It is not surprising that eigenvalues at 
the edge of the block have larger errors.
Other large errors could be reduced by increasing $r$. 
Some of the eigenvectors  corresponding to eigenvalues  near the top of the block are nearly linear 
combinations of the eigenvectors  {\it{ of different symmetries}} 
computed with the Smolyak quadrature method of  Ref. \onlinecite{avila2011}. 
We  have not yet implemented a symmetry-adapted rank reduction and therefore the  rank reduction 
breaks    the symmetry.   
In some cases, when levels are close together,  the RRBPM energies and those of  Ref. \onlinecite{avila2011}
are not  close enough to match them unambiguously.   When this problem occurs we match levels using their
assignments, i.e., using the corresponding eigenvectors.  
Differences between the energies of ref. \onlinecite{avila2011} and experiment are due to the potential.    
Differences between the energies of ref. \onlinecite{avila2011} and those obtained with the 
RRBPM are due 
to the low-rank approximations and
 could be reduced by increasing $r$, $B$, the maximum $N_{\text{pow}}$ and $N_{ALS}$. 

\begin{table*}[ht]  
\caption{Transition wavenumbers from the ZPE.  From left to right:  RRBPM  results, symmetry, 
assignment, results from references \onlinecite{avila2011} and \onlinecite{begue2005}, experimental values. 
The zero point energy is 9837.6293 cm$^{-1}$. 
RRBPM energies  and the energies of Ref. \onlinecite{avila2011} are matched on the basis of assignments.  The
symmetry is taken from  Avila et al\cite{avila2011}.
States  in a curly bracket are linear combinations (LC) of states in Ref. \onlinecite{avila2011}.
 }
\begin{tabular}{cccccc}
\hline
\hline
Transitions (cm$^{-1}$) & Sym. & Assign. & Ref. \onlinecite{avila2011} & Ref. \onlinecite{begue2005}  & Exp.  \\
\hline
361.18, 361.25  & E & $\omega_{11}$ & 360.991  & 366  & 362 (Ref. \onlinecite{exp108}), 365 (Ref. \onlinecite{exp109})  \\
723.37, 724.38  & E & $2\omega_{11}$ & 723.181  & 725  & 717 (Ref. \onlinecite{exp108} and \onlinecite{exp110}) \\
724.96 & A$_1$ & $2\omega_{11}$ & 723.827  & 731  & 739 (Ref. \onlinecite{exp110})\\
900.97 & A$_1$ & $\omega_4$ & 900.662  & 916  & 916 (Ref \onlinecite{exp108}), 920 (Ref. \onlinecite{exp109}) \\
1034.50, 1034.55  & E & $\omega_9$ & 1034.126  & 1038  & 1041 (Ref \onlinecite{exp108}), 1042 (Ref. \onlinecite{exp110}) \\
1087.95 & A$_2$ & $3\omega_{11}$ & 1086.554  & 1098  & 1122 (Ref. \onlinecite{exp110}) \\
1088.58 & A$_1$ & $3\omega_{11}$ & 1086.554  & 1098  & 1122 (Ref. \onlinecite{exp110})\\
1090.75, 1090.85  & E & $3\omega_{11}$ & 1087.776  & 1094  & 1077 (Ref. \onlinecite{exp110})\\
1260.89, 1261.12  & E & $\omega_4+\omega_{11}$ & 1259.882  & 1282  & 1290 (Ref. \onlinecite{exp111}) \\
1391.76 & A$_1$ & $\omega_3$ & 1388.973  & 1400  & 1390 (Ref \onlinecite{exp112}), 1385 (Ref. \onlinecite{exp111})\\
\multirow{2}{*}{
$
\left.
\begin{array}{rcl}
1395.74, 1398.24 \\
1396.24
\end{array}
\right\rbrace
$
}
& \multirow{2}{3cm}{L.~C. of E and A$_2$ states} & $\omega_9+\omega_{11}$ & 1394.689 (E)& 1401 & 1410 (Ref. \onlinecite{exp112}), 1409 (Ref. \onlinecite{exp111})\\
  &  & $\omega_9+\omega_{11}$ &  1394.907 (A$_2$) &  1398 (A$_2$) & 1402 (Ref. \onlinecite{exp111})\\
1401.15 & A$_1$ & $\omega_9+\omega_{11}$ & 1397.687  & 1398  & 1402 (Ref. \onlinecite{exp111})\\
1452.92, 1458.62  & E & $4\omega_{11}$ & 1451.101  &   &  \\
\multirow{2}{*}{
$
\left.
\begin{array}{rcl}
1456.24, 1460.80 \\
1464.40
\end{array}
\right\rbrace
$
}
 &\multirow{2}{3cm}{L.~C. of E and A$_1$ states} & $4\omega_{11}$ & 1452.827 (E) & & \\
  &  & $4\omega_{11}$ &   1453.403 (A$_1$) & 1467  & 1448 (Ref. \onlinecite{exp111})\\
1483.52, 1483.54  & E & $\omega_7$ & 1483.229  & 1478  & 1453 (Ref. \onlinecite{exp108}), 1450 (Ref. \onlinecite{exp111})\\
1621.34, 1623.05  & E & $\omega_4 + 2\omega_{11}$ & 1620.222  & 1647.0  &  \\
1624.05 & A$_1$ & $\omega_4 + 2\omega_{11}$ & 1620.767  & 1645.7  &  \\
1753.66, 1755.03  & E & $\omega_3 + \omega_{11}$ & 1749.530  & 1766.4  &  \\
1759.62, 1760.16  & E & $\omega_9+2\omega_{11}$ & 1757.133  & 1767.4  &  \\
1765.59 & A$_1$ & $\omega_9+2\omega_{11}$ & 1756.426  & 1761.6  &  \\
1779.88 &  A$_2$ & $\omega_9+2\omega_{11}$ & 1756.426  & 1761.6  &  \\
1780.66, 1780.86  & E & $\omega_9+2\omega_{11}$ & 1759.772  & 1769.3  &  \\
1786.17 & A$_1$ & $2\omega_4$ & 1785.207  & 1833.7  &  \\
1823.34, 1830.31  & E & $5\omega_{11}$ & 1816.799  &   &  \\
1823.87, 1828.40  & E & $5\omega_{11}$ & 1820.031  &   &  \\
1827.34 & A$_2$ & $5\omega_{11}$ & 1818.953  &   &  \\
1832.19 & A$_1$ & $5\omega_{11}$ & 1818.952  &   &  \\
\multirow{2}{*}{
$
\left.
\begin{array}{rcl}
1845.57 \\
1846.85, 1849.44
\end{array}
\right\rbrace
$
}
 & \multirow{2}{3cm}{L.~C. of E and A$_2$ states} & $\omega_7+\omega_{11}$ & 1844.258 (A$_2$), &  1838.2 (A$_2$), & \\
 & & $\omega_7+\omega_{11}$ &  1844.330 (E) & 1842.2 (E) &  \\
1848.14 & A$_1$ & $\omega_7+\omega_{11}$ & 1844.690  & 1838.2  &  \\
1932.98, 1933.43  & E & $\omega_4 + \omega_9$ & 1931.547  & 1952.3  &  \\
1990.80, 1992.59  & E & $\omega_4 + 3\omega_{11}$ & 1982.857  & 2015.1  &  \\
2000.43 & A$_2$ & $\omega_4 + 3\omega_{11}$ & 1981.850  & 2010.3  &  \\
2007.90 & A$_1$ & $\omega_4 + 3\omega_{11}$ & 1981.849  & 2010.3  &  \\
2058.94 & A$_1$ & $2\omega_9$ & 2057.068  & 2059.0  & 2075 (Ref. \onlinecite{exp108})\\
2066.43, 2068.26  & E & $2\omega_9$ & 2065.286  & 2067.0  & 2082 (Ref. \onlinecite{exp108})\\
\multirow{2}{*}{
$
\left.
\begin{array}{rcl}
2116.66, 2121.90 \\
2136.72
\end{array}
\right\rbrace
$
}
 &\multirow{2}{3cm}{L.~C. of E and A$_1$ states}	&	$\omega_3 +2\omega_{11}$	&	2111.380 (E)& 2131.3 (E)	& \\ 
 & & $\omega_3 +2\omega_{11}$ & 2112.297 (A$_1$) & 2130.0 (A$_1$) & \\
\multirow{6}{*}{
$
\left.
\begin{array}{rcl}
2126.04 \\
2144.72 \\
2146.06 \\
2174.61 \\
2150.37 \\
2153.59 
\end{array} 
\right\rbrace
$
}
	& \multirow{6}{3cm}{L.~C. for these six states. Two levels are missing.	} &	$\omega_9 + 3 \omega_{11}$	&	2119.327 (E) & &\\
	&		& $\omega_9 + 3 \omega_{11}$ &	2120.541 (E) & &\\
	&		& $\omega_9 + 3 \omega_{11}$ &	2120.910 (A$_2$) & &\\
	&		& $\omega_9 + 3 \omega_{11}$ &	2122.834 (E) & &\\
	&		& $\omega_9 + 3 \omega_{11}$ &	2123.301 (A$_1$) & &\\
	&		& $\omega_9 + 3 \omega_{11}$ &			& & \\
2158.75, 2164.96 &	E	&	$2\omega_4+ \omega_11$	& 2142.614 & 2199.4	& \\
2210.70	& ? &	$6\omega_{11}$	& 2183.635 (E) &  & \\
\hline
\hline
\end{tabular}
\label{resultsCH3CN}
\end{table*}

\section{Conclusion}

The use of iterative algorithms has opened the door to routine calculation of (ro-)vibrational spectra of 
molecules with as many as four or five atoms.\cite{mjbdirect,corey,dvrrev,lef,lef2,cg,crunch,muck}
The same ideas can be used, with an adiabatic approximation, for Van der Waals complexes with 6 or fewer 
inter-molecular coordinates.\cite{he2,nno}
Although iterative methods obviate the need to  store the Hamiltonian matrix 
(or even to calculate its matrix elements),  application of these ideas to larger molecules is impeded by the size of 
the vectors that must be stored in memory.   Calculations are only ``routine'' if a product basis is used. 
With a product basis, the size of a vector scales as $n^D$.  For a $J=0$ calculation, $D= 12$ for  molecule with 6 atoms;
storing a vector with $10^{12}$   elements requires 8000 GB.   One way to deal with this impasse is use a 
contracted basis.   Another is to prune a product basis set.   In this article we suggest a 
third approach, based on exploiting the SOP form of the Hamiltonian.   At present it can only be applied
to SOP Hamiltonians.   That is a limitation,  but for many molecules with 6 or more atoms for which one wishes 
to compute a spectrum  either the only available potential energy surfaces (PESs) are in SOP form or the PES can be brought into SOP form 
without making a significant approximation.  MCTDH is usually used with a SOP PES, there are however other options.
\cite{uwe1,uwe2}

The principal idea of the RRBPM  of this paper is the realization that whereas one needs $n^D$ numbers to 
represent a general function in a product basis, a function that is a SOP can be represented with far 
fewer numbers.  The simplest example is a product of $D$ factors for which one only requires $nD$ numbers,
much less than $n^D$.   If the factors of the terms in a SOP representation of a wavefunction are 
chosen carefully, it should be possible to represent it with a relatively small number of terms.  In this article,
we show that it is possible to obtain accurate energy levels for a 6-D model problem with $r=10$,  for a
20-D model problem with $r=20$ and for   CH$_3$CN  (12-D) with 299 
coupling terms with   $r=20$.  This makes it possible to reduce the memory cost of calculations by 
many orders of magnitude. 
For the 20-D problem the memory cost is about  1 GB.  
 The RRBPM uses a shifted block power method to make SOP basis functions.  Applying the 
Hamiltonian to a vector necessarily yields the number of terms in the SOP.  If this increase were not 
checked the memory cost of the method would become large.  We restrict the number of terms by using a 
rank reduction idea.  \cite{beylkin2005}
A somewhat similar power method idea has been used in  Ref. \onlinecite{nakatsuji}.  
At each stage of the procedure we exploit the SOP structure, e.g., to orthogonalize, to evaluate matrix-vector
products etc.

The ideas introduced in this paper can be refined in several ways.   Rather than using the block power
method to generate SOP functions one could use a better iterative approach.  One option is a 
 Davidson algorithm\cite{davidson1975}, another is a preconditioned inexact spectral transform 
(PIST) method.\cite{pist1,pist2,billpist}  
A PIST version would also make it possible to target high-lying levels.  
Reducing the number of required iterations would reduce the cost of the calculations.
Any iterative method whose basis vectors are close enough to the desired 
 eigenvectors to enable rank reduction will be suitable.
A symmetry-adapted rank reduction method will obviate symmetry mixing of very nearly degenerate levels.   
The  ALS reduction algorithm is not the most efficient nor the most robust reduction algorithm in the literature.   
It could be replaced, for example, with 
 a conjugate gradient-based algorithm\cite{espig2012}. 
The general approach is promising because its memory cost is low.      
It might be possible to use similar ideas with the Floquet formalism to solve  the 
 time-dependent Schr\"odinger equation to study a molecule in 
a  strong external electromagnetic 
field.\cite{chu2003} 
In  such approaches memory cost is a serious problem because  the time-dependent Schr\"odinger equation 
is solved in  an extended Hilbert space containing functions that depend on a  time coordinate.\cite{leclerc2011,leclerc2012} 
We have shown that the method can be used to solve      the time-independent Schr\"odinger equation for  a molecule with 
6 atoms using less  than 1 GB.    Clearly, it will be  possible to compute spectra for much larger molecules.

\begin{acknowledgments}
We  thank  Gustavo Avila for his help. 
Calculations  have been executed on computers of  the Utinam Institute at
 the Universit\'e de Franche-Comt\'e, supported by the R\'egion de Franche-Comt\'e and Institut des Sciences de l'Univers (INSU) and on computers purchased with a grant for the Canada Foundation for Innovation.  This research was funded by the 
Natural Sciences and Engineering Research Council of Canada.  We thank James Brown for doing the 6-D direct-product
Lanczos calculation.

\end{acknowledgments}



%

\end{document}